\documentclass[pre,aps,12pt]{revtex4-2}
\pdfoutput=1
\usepackage{amsmath}
\usepackage{amssymb}
\usepackage{physics}
\usepackage{hyperref}
\usepackage{color}
\usepackage{bbm}
\usepackage{graphicx}% Include figure files
\usepackage{dcolumn}% Align table columns on decimal point
\usepackage{bm}% bold math
\usepackage{comment}

\usepackage{tikz}
\usepackage{wasysym}
\usetikzlibrary{fit}
\usepackage{nicefrac} 

\usepackage{tikz-network}
\usetikzlibrary{patterns,decorations.pathreplacing}
\usepackage{tabularx}
\usepackage{tabu}
\usepackage{multirow}
\definecolor{myred}{RGB}{232,102,102}
\definecolor{myblue}{RGB}{150,150,255}
\definecolor{myorange}{RGB}{255,165,0}
\definecolor{mygrey}{RGB}{105,105,105}
\definecolor{OliveGreen}{RGB}{85,107,47}
\definecolor{NavyBlue}{RGB}{0,0,128}
\definecolor{mygreen}{RGB}{34,139,34}
\definecolor{myY}{RGB}{220,255,203}
\definecolor{myYO}{RGB}{255, 220, 151}
\definecolor{color1}{RGB}{230,230,250}
\definecolor{color2}{RGB}{255,255,224}
\definecolor{green2}{RGB}{0,100,0}

\usepackage{mathtools}

\newcommand{\Wgateteal}[2]{
\draw[very thick] (#1-0.5, #2 +0.5) -- (#1+0.5,#2-0.5);
\draw[very thick] (#1-0.5,#2-0.5) -- (#1+0.5,#2+0.5);
\draw[ thick, fill=teal, rounded corners=0.8pt] (#1-0.25,#2+0.25) rectangle (#1+0.25,#2-0.25);
\draw[thick] (#1,#2+0.15) -- (#1+0.15,#2+0.15) -- (#1+0.15,#2);
}
\newcommand{\newc}{\newcommand}
\newc{\beq}{\begin{equation}}
\newc{\eeq}{\end{equation}}
\newc{\kt}{\rangle}
\newc{\br}{\langle}
\newc{\fm}{\fullmoon}
\newc{\hK}{\hat{\mathcal{K}}}
\newc{\mP}{\mathcal{\phi}}
\newc{\mX}{\mathcal{X}}
\newc{\discuss}[1]{\textcolor{red}{#1}}

\begin{document}

%\preprint{APS/123-QED}

\title{Quantum-classical correspondence in quantum channels}% Force line breaks with \\

\author{Bidhi Vijaywargia}
\email{bidhi.vj08@gmail.com}
\affiliation{%
Department of Physics, \& Center for Quantum Information, Computation and Communication, \\ Indian Institute of Technology Madras, Chennai, India~600036}
\author{Arul Lakshminarayan}%
 \email{arul@physics.iitm.ac.in}
 \affiliation{%
Department of Physics, \& Center for Quantum Information, Computation and Communication, \\ Indian Institute of Technology Madras, Chennai, India~600036}

\date{\today}% It is always \today, today,

\begin{abstract}
Quantum channels describe subsystem or open system evolution. Using the classical Koopman operator that evolves functions on phase space, 4 classical Koopman channels are identified that are analogs of the 4 possible quantum channels in a bipartite setting. Thus when the complete evolution has a quantum-classical correspondence the correspondence at the level of the subunitary channels can be studied. The channels, both classical and quantum can be interpreted  as noisy single particle systems. Having parallel classical and quantum operators gives us new access 
to study fine details of these major limiting theories. Using a coupled kicked rotor as a generic example, we contrast and compare spectra of the quantum and classical channel. The largest nontrivial mode of the quantum channel is seen to be mostly determined by the stable parts of the classical phase space, even those that are surprisingly small in relation to the scale of an effective $\hbar$. In the case when the dynamics has a significant fraction of chaos the spectrum has a prominent annular density that is approximately described by the single-ring theorem of random matrix theory, and the ring shrinks in size when the classical limit is approached. However, the eigenvalues and modes that survive the classical limit seem to be either scarred by unstable manifolds or, if they exist, stable periodic orbits.

\end{abstract}

\maketitle

%\tableofcontents

\section{\label{sec:intro}Introduction}

Quantum chaos in closed single particle systems has been studied extensively, for example see the textbooks \cite{Reichl2004,haake2001quantum,gutzwiller1990chaos}. The approaches include statistical analysis, mainly using random matrix theory (RMT) \cite{haake2001quantum} or semiclassical ones such as Gutzwiller's periodic orbit theory \cite{Gutzwiller1971semiclassical}. One of the major themes has been quantum-classical correspondence as classical chaos in phase space arising from nonlinear dynamics is seen in contrast with the linear quantum evolution in Hilbert space. Thus many ``signatures" of classical chaos in quantum systems have been found. However, even without the complications of chaos, quantum-classical correspondence is subtle as the $\hbar \rightarrow 0$ limit is singular (here $\hbar$ is a dimensionless scaled and effective Planck constant). In such limiting theories the asymptotic regime can lead to new physics such as in turbulence and critical phenomena, as argued by Berry in \cite{BerryLesHouches}. More recently, quantum chaos in closed many-body systems is being vigorously pursued from many different perspectives such as thermalization \cite{ETH_review}. In most of these, the classical limit is either nonexistent or at least not obvious, such as in fermionic systems or spin-1/2 ``qubits". Further, nonclassical phenomena such as entanglement play an important role in the characterization and description of many-body chaos, for example \cite{Kim13,Kolovsky16,Kaufman16}.

Dissipative, non-Hermitian or open quantum systems and especially the development of quantum chaos havs been of continuing interest from early on \cite{GrobePRL88,Brun_1996}. It has seen a recent spurt of activity, from extending the Bohigas-Giannoni-Schmidt conjecture, linking RMT to quantization of chaotic systems, to the elucidation of a 38 fold symmetry classification \cite{Xu19,Akemann19,Lucas2020,Garcia22,Kawabata23}. However, there have been several approaches to quantum chaos in open systems.
 In one case a part of the classical phase space was excised to create the open or dissipative dynamics. Thus the quantum propagator involves a projection operator following the unitary one. Spectra and resonances of such quantum open systems have been quite extensively studied, for example see \cite{Keating2006,Wisniacki_openscar2008}.  
Another approach is to consder effective complex Hamiltonians or introduce some other forms of non-Hermiticity, PT symmetric or otherwise \cite{Manas_nonherm,Mudute-Ndumbe_2020nonherm,Jisha_2024nonherm,Joseph_nonherm}. The Lindblad approach provides a master equation for the evolution of the density matrix when the system weakly interacts with a Markovian environment. The spectrum of the Liouvillian or its exponential is of interest in this context. Recent works have constructed and examined properties of random Lindblad and Liouvillian operators \cite{Can_2019lindblad,Alvaro_liouvillian2022,Wojceich_2019lindblad,Lucas_Prosen_singlering}.

The approach in this paper is bipartite and exact, in the sense that we consider channels that are usually described by tracing out the environment. These channels are completely positive trace preserving (CPTP) maps and take density operators to other density operators. We consider cases where the complete system+environment has a unitary evolution and has a well-defined classical limit. We then use the unitary classical Koopman operators on this combined space to define classical channels in a manner completely analogous to the quantum. There are four possible channels and we study in depth one of them in both its classical and quantal manifestations. We show explicitly how the act of tracing out one degree of freedom results in the classical Koopman channel's interpretation as stochastic Koopman operator, of a noisy one-degree of freedom system. Similarly we show that the quantum channel is an example of a dephasing Hadamard channel.

The quantum channels and the classical channels are two objects in different theories, but they are most suited for comparisons and in the long history of the correspondence principle, this seems to provide a nonsingular limiting process, unlike many of the usual semiclassical aymptotics plagued by convergence issues. We consider in some detail a two degree of freedom system, coupled kicked rotors or standard maps, each of which can exhibit chaos on their  own. The standard map provides a convenient platform as it is a well understood (at least studied) system with a typical Hamiltonian scenario for the development of chaos, and an integrable-nonintegrable transition via the KAM (Kolmogorov-Arnold-Moser) tori breaking into cantori and losing stability. The map has been studied here in the toral setting, but on the cylinder has long been implemented experimentally using cold atoms \cite{Raizen1994}.

 In this work, we would like to understand how the eigenvalues are distributed inside the unit disk and what is the effect of the classical dynamics on the eigenvalues of the quantum channels. In the quantum case we find two kinds of eigenvalues, those which cluster inside a more-or-less well defined annular region, a ``single ring", and those which are outside of it. The ring contracts with effective Planck constant, in apparent agreement with the single-ring theorem of RMT \cite{FEINBERG_ZEE,Guionnet2011}, and seems dominated by the chaotic or random states, while those outside seem to survive the classical limit and 
 are dominated by imprints of regular islands if any, or unstable manifolds. 
 The spectra of the quantum channels are thus susceptible to the happenings in classical phase space and the presence of tiny islands of regularity can take over the largest eigenmodes and determine the decay rates of correlations.

 Most importantly perhaps, we have truly analogous classical and quantum objects to compare. There are subunitary operators in both domains, and we can compare both eigenstates and eigenvalues directly. Thus the phenomenon of quantum scarring of eigenstates we also find in the classical, especially by the presence of classical unstable manifolds. The effects of noise or interaction on these phenomenon can be directly and simply studied. 
Although the setting is bipartite, the motivations come from unitary circuits built from such bipartite systems, for example in brick-wall configurations \cite{Dual_Prosen,Dual_Koopman_Arul,Dual_sympletic_Prosen}. There are implications for such many-body systems, from the decay of two-body correlations to perhaps many-body scarring propagating from the scarring of these channels \cite{Logari2024}.

\subsection{\label{subsec:motiv} Some background and motivation}

The quantization of a symplectic classical dynamical map is associated with a unitary operator, $U$ \cite{BERRY1979, BALAZS19891, N.L.Balazs_1987, Baker_Saraceno, Reichl2004,haake2001quantum,gutzwiller1990chaos}, the corresponding quantum map. It has been known for quite some time now that the level statistics obtained from the eigenangles of the quantum map, the quasi-energies of $U$, show universal behaviour in classically chaotic regimes\cite{casati1985chaotic,IZRAILEV1990299}. Furthermore, the eigenstates of the quantum map can carry information about the phase-space structures of the associated classical map, such as stable and unstable periodic orbits and homoclinic orbits \cite{Berry1989_scar, Heller, Baker_Saraceno}.

Consider, for example, the two-dimensional area-preserving standard map \cite{lichtenberg1983regular, Reichl2004, chirikov1981dynamical}: 
    \begin{equation}
      \label{eq:stand_map}
    \begin{split}
            q' & = q + p' \\
            p' & = p - (\alpha/2\pi)\sin(2\pi q)
            \end{split}
    \end{equation}
The physical system resulting in the above transformation is a pendulum that is being kicked periodically with a time period $T=1$, with an impulsive force $=-\alpha \sin (2 \pi q)/2 \pi$. The motion can be restricted to a cylinder or torus by applying the modulo-$1$ operation to the first or both of the above equations.

The corresponding quantum map, or Floquet operator, is given by  
\begin{equation}
    \label{eq:quan_stand_map}
    U = \exp\left(\frac{- ip^{2}}{2\hbar}\right) \exp \left( \frac{i\alpha}{4\pi^{2} \hbar} \cos(2\pi q)\right)
\end{equation}
Restricting the dynamics to the unit torus, with periodic boundary conditions of unit period in both $q$ and $p$ directions, results in a finite Hilbert space with dimension $N = 1/h$ and the classical limit is $N \rightarrow \infty$ \cite{IZRAILEV1990299,CHIRIKOV198877}.
The eigenspectra are defined by 
\begin{equation}
\label{eq:eig_unit_stand}
U|\psi_{j}\kt = \exp(i\psi_{j})|\psi_{j}\kt \,\,\, , 0 \leq j \leq N-1
\end{equation}

The Husimi distribution, which is the intensity of the coherent state representation has been useful in identifying classical phase space structures in quantum states \cite{Takahashi1985,husimi1940density, Baker_Saraceno}, and is defined as
%\begin{equation}
%    \label{eq:hus_unit_state}
    $h_{\psi}(p,q) = |\br pq| \psi \kt |^{2}$
%\end{equation}
where, $|pq\kt$ is a minimum uncertainty coherent state centered at $(q,p)$ on the phase space. 
\begin{figure}[h]
\includegraphics[scale=.5]{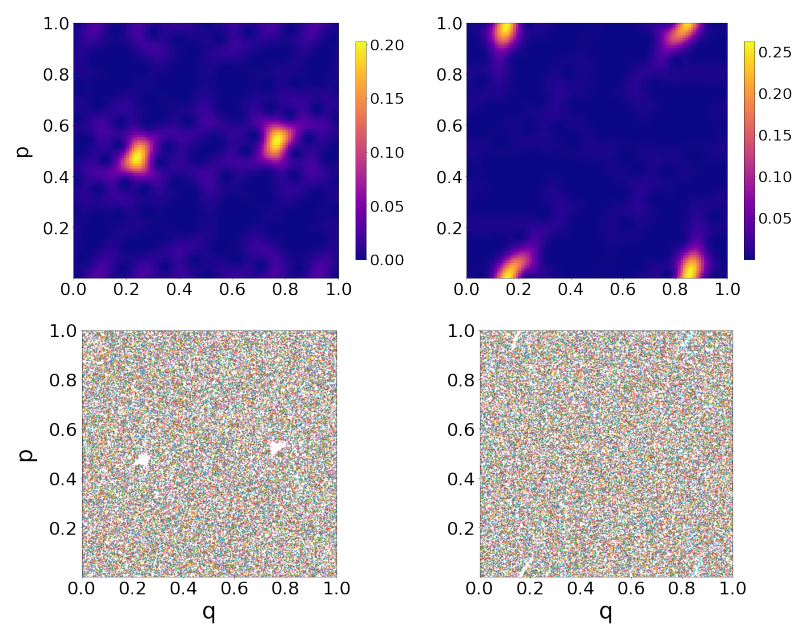}
\caption{\label{fig:hus_unit_state} The Husimi representation of two eigenstates of the Floquet operator for the standard map alongside the classical phase space corresponding to $\alpha=6.0$, and $\alpha=7.5$ for the case $N=100$.  }%
\end{figure}
For example, classical stable orbits or tori in regular regions and unstable periodic orbits in chaotic regions can show up as features in the Husimi functions of eigenstates, the latter are referred to as ``scars" \cite{Heller} and were not initially expected as they are of measure $0$, even if dense. 
In fact, less studied is when the regular islands are being squeezed out of existence when they are loosing stability. Somewhat surprisingly, even tiny islands of regularity, maybe smaller than the effective $h$, can influence quantum eigenstates. Two examples are  shown in Fig.~(\ref{fig:hus_unit_state}). Here the first case at $\alpha=6.0$ is a period-2 island that is somewhat smaller in area than $1/N=.01$ and in the second case are tiny remnants of islands corresponding to an accelerator mode for the map on the cylinder \cite{CHIRIKOV1979263,lichtenberg1983regular}, and are barely visible.
A study of ``superscars" at bifurcations exists, where the analysis is of an average over a semiclassially large number of states \cite{Keating_superscar}, also see \cite{BOHIGAS1993}. While we do not study this phenomenon directly, such states do show up gratuitously on the inclusion of noise, which we study below.

There is no direct classical analogue of the unitary propagator, although there are semiclassical approximations \cite{vanvleck1928correspondence,Gutzwiller1971semiclassical}. However, consider the ``adjoint action" in the Heisenberg picture of quantum mechanics: $\hat{A'}=U^{\dagger} \hat{A} U$ which evolves an operator $\hat{A}$
(or the evolution of a density matrix $\hat{\rho}'=U \hat{\rho} U^{\dagger}$ in the Schr\"odinger picture). Let $\bra{i} \hat{A} \ket{j}\equiv \bra{ij}\ket{\hat{A}}$ be the ``row-wise" vectorization of the operator $\hat{A}$. Noting the identity $\ket{\hat{A}\hat{B} \hat{C}}=(\hat{A}\otimes \hat{C}^T)\ket{\hat{B}}$, where $\cdot^T$ denotes transpose, we have that $\ket{\hat{A}'}=(U^{\dagger} \otimes U^{T})\ket{\hat{A}}$ and $\ket{\hat{\rho}'}=(U \otimes U^*) \ket{\hat{\rho}}$. The classical analogues of these actions may be considered to be the Koopman operator \cite{Koopman}
and its adjoint, the Perron-Frobenius.  If $\Omega$ denotes the phase space and $\Phi_t: \Omega \mapsto \Omega$ is a dynamical system, and the subscript $t$ denotes time. The Koopman operator is defined by its action on functions (observables) $f(x)$ as simply the composition operator:
\begin{equation}
    \label{eq:koop}
    \mathcal{\hat{K}}_tf(x) = f[\Phi_t(x)]=\int_{\Omega} \delta[y-\Phi_t(x)] f(y) dy=\int_{\Omega} \mathcal{K}_t(x,y)f(y) dy,
\end{equation}
where the kernel $\mathcal{K}_t(x,y)=\delta[y-\Phi_t(x)]$ is the Koopman propagator ``matrix elements". 
We are interested in the case when $\Phi_t$ is a canonical transformation generated by a Hamiltonian. If the classical phase space is represented by  $\xi = (q,p)$, and $F_n(\xi)$
is a set of orthogonal functions on it (say with the $L^2$ inner-product, as in a Hilbert space), the Koopman operator elements in the Dirac bra-ket notation may be expressed in the following forms, the latter being restricted to the space spanned by the orthogonal set: 
\begin{equation}
    \label{eq:koopman}
    \br \xi |\hat{\mathcal{K}_t}|\xi' \kt = \delta [\xi'-\Phi_t(\xi)], \;\; \br F_{n'}|\hat{\mathcal{K}}_t|F_n \kt= \int_{\Omega} F^*_{n'}(\xi) F_n[\Phi_t(\xi)] d \xi.
\end{equation}
This is an infinite dimensional unitary matrix and hence the curse of dimensionality is the prize to pay for the linearization. More mathematically rigorous statements concerning the nature of the spaces of functions and proofs  can be found in references such as \cite{Koopman,Gaspard_1998,Mezic_koopbook}. Note that here we have taken $f(\xi)=\bra{\xi}\ket{f}$ and as
$\bra{\xi}\hK_t \ket{f}=\bra{\Phi_t(\xi)}\ket{f}$,
if we isolate the action on the dual vector space $\bra{\xi}$, we have $\bra{\xi}\hK_t=\bra{\Phi_t(\xi)}$. We can think of $\ket{\xi}$ as representing the density $\rho(\xi')=\bra{\xi'}\ket{\xi}=\delta(\xi'-\xi)$, and $\bra{\xi'}\hK_t^{\dagger} \ket{\xi}=\delta[ \xi'-\Phi_t(\xi)]$ is the adjoint of the Koopman, the Perron-Frobenius operator. Its action on general densities on the phase space is $\bra{\xi} \hK_t^{\dagger} \ket{\rho}=\rho[ \Phi^{-t}(\xi)]$, where we have assumed reversible flows as Hamiltonian flows are. The Koopman {\em vs }Perron-Frobenius duality is the classical shadow of the Heisenberg {\em vs} Schr\"odinger pictures of quantum mechanics.

Note that for any quantum operators $\hat{A}$ a psuedo phase space function can be defined via the Wigner transform. Therefore the evolution operator may be sought as a direct rather than an adjoint action, and hence the Koopman is more an analogue of the unitary channel $U^{\dagger} \otimes U^T$ rather than simply the propagator over time $t$, $U$ itself. In other words the observables of classical mechanics and quantum mechanics are treated to the direct unitary actions, and we must consider the vector space of quantum operators rather than states. This Hilbert space formulation of classical mechanics implies that all classical mechanics is a linear unitary evolution, just as quantum mechanics is. Thus,  discussions of quantum {\em vs} classical chaos that rest on the linear {\em vs} nonlinear aspects are more or less red herrings. The Koopman formalism has gained much attention  recently in the context of nonlinear control and beyond \cite{KORDA2018149koopmancontrol,brunton2016koopmancontrol,Kaiser_2021koopmancontrol,PEITZ2019184koopmancontrol,kaiser2018sparsekoopmancontrol,shi2022deepkoopmanoperatorcontrol,yeung2017learningdeepneuralnetworkcontrol,lusch2018deepcontrol}. 

In Sec.~\ref{sec:channels} both the classical Koopman and quantum 
channels are introduced and  interpreted as a stochastic Koopman operator and as a Hadamard dephasing channel respectively. In Sec.~\ref{sec:spectra} we introduced the details of the coupled kicked rotors which are studied in detail. The quantum and classical channel eigenvalues are compared with each other as a function of effective Planck constant, as a function of the kicking strength governing the extent of chaos and of the noise strength. We also point out the relevance of entangling power of the whole unitary in determining some gross features such as the outer radius of an annular ring on which most eigenvalues cluster. Thus we make connections to one of RMT's celebrated results, the single-ring theorem.
In Sec.~\ref{sec:Modes} we display some of many modes of both the channels, the quantum ones in the Husimi representation. We find intriguing structures in both as the principal eigenvalues pick out the localized states such as scarred ones and lie outside the random single-ring region. We also discuss the possible presence of chaos assisted tunneling in such channels, and its interplay with noise. We also present an approach that is numerically viable for small couplings that separates out the many localized significant modes from the others. Finally we summarize and conclude in Sec.~\ref{sec:Summary}.

\section{\label{sec:channels} The Channels and their properties}%In this section, we briefly introduce the concept of quantum channels. 
\subsection{\label{subsec:quan_channel} Quantum Channels}

Consider a bipartite unitary evolution $U$ acting on states from a product Hilbert space $\mathcal{H}_1^N\otimes \mathcal{H}_2^N$. We will consider the case when the subsystems have equal dimensionality of $N$, the generalizations are not hard. The Heisenberg evolution of a local operator in subsystem $1$ is $U^{\dagger}\left(\hat{a} \otimes \mathbbm{1}\right)U$. This is typically an entangled (``scrambled") operator on the product space. It is useful to define a partial traced local operator as
\begin{equation}
\label{eq:opt_evolve}
\hat{a}'=\text{Tr}_{2}\left[U^{\dagger}\left(\hat{a} \otimes \frac{\mathbbm{1}}{N}\right)U\right].
\end{equation}
The initial operator on subsystem $2$ is scaled so that it has unit trace, as well as if we were to think of action on states, this would correspond to the maximally mixed environment. This also ensures that the $\Tr_1 \hat{a}'=\Tr_1 \hat{a}$. A quantum channel connects the two subsystem operators $\hat{a}$ and $\hat{a}'$ and is best expressed in terms of their vectorizations (see above) as
\begin{equation}
\label{eq:qchan_act}
   |\hat{a}'\kt = \mathcal{M}(U)|\hat{a}\kt, \;\; \text{where} \;\;  
    \mathcal{M}(U)= \frac{1}{N}[U^{R_{1}\dagger}U^{R_{1}}]^{R_{2}}
\end{equation}
The dependence of the channel only on the evolution $U$ is emphasized in the notation. Note that $\mathcal{M}(U)$ is a $N^{2} \times N^{2}$ shaped matrix that is in general sub-unitary. Here, $R_{i}$ denotes realignment operations, which are defined as 
    $\langle i \alpha|U|j \beta \rangle = \langle \beta\alpha |U^{R_{1}}|ji\rangle =     \langle i j |U^{R_{2}}|\alpha \beta \rangle$. Indeed these channels are examples of completely positive trace preserving (CPTP) operators. In this explicit form they are written as a realigment of a positive operator \cite{ECG_CP}. 
    
Note that in general the realignment $U^{R_1}$ is itself not unitary. The class of $U$ for which $U^{R_1}$ (equivalently $U^{R_2}$) is unitary have been recently referred to as ``dual unitary" \cite{Dual_Prosen} and are of maximal operator entanglement \cite{Suhail_EnsembleDU}. If $U$ is dual unitary, then $\mathcal{M}(U)=\mathbbm{1}_{N^2}^{R_2}/N=\ketbra{\fullmoon}{\fullmoon}$, where $\ket{\fm} = \sum_{k=1}^N \ket{kk}/\sqrt{N}$ is the standard maximally entangled state, proportional to the vectorization of the $N$ dimensional identity matrix. Therefore 
$\ket{\hat{a}'}=\Tr(\hat{a}) \ket{\fm}/\sqrt{N}$, or 
$\hat{a}'= \mathbbm{1}_N \Tr(\hat{a})/N $. The same considerations hold for evolution of states and in this case {\em any} normalized state $\rho$ is replaced by the most mixed state on substsem $1$, in other words the channel $\mathcal{M}(U)$ is a completely depolarizing one. Partial traced operators are useful to define two-point dynamical correlations such as
\begin{equation}
\label{eq:corr_fold}
    D_Q^{(11)} = \text{Tr}\left[\left(\hat{a}_0 \otimes  \mathbbm{1}\right) U^{\dagger} \left(\hat{a} \otimes \frac{\mathbbm{1}}{N} \right)U \right] = \text{Tr}_{1}(\hat{a}_0\hat{a}')= \bra{\hat{a}_0^{\dagger}}\ket{\hat{a}'}= \bra{\hat{a}_0^{\dagger}}\mathcal{M}(U) \ket{\hat{a}}.
\end{equation}
If $U$ is dual unitary then $D_Q^{(11)}=\Tr_1(\hat{a}_0^{\dagger})\Tr_1(\hat{a})/N$, indicating complete operator mixing. Thus the spectrum of the channel $\mathcal{M}(U)$ determines this correlation function. However, their meanings as decay rates are not obvious as there is no multiplicative property, in particular $\mathcal{M}(U^2)\neq \mathcal{M}^2(U)$. In fact even for dual unitaries this is not true as powers do not respect such duality. 

The complementary channel \cite{Holeve_compchann,Smaczy_compchann,Devetak2003complementarychannel} associated with $\mathcal{M}(U)$ is defined as 
\begin{equation}
    \label{eq:comp_qchan}
    |\hat{b}'\kt = \mathcal{M}_{C}(U)|\hat{a}\kt, \;\; \text{where}\;\; \hat{b}' = \text{Tr}_{1}\left[ U^{\dagger} \left(\hat{a} \otimes \frac{\mathbbm{1}}{N}\right) U \right].
\end{equation}
This channel can be written in terms of the partial transpose of $U$: 
\begin{equation}
    \label{eq:form_comp_qchan}
     \mathcal{M}_{C}(U) = \frac{1}{N}[U^{T_{2}}U^{T_{2}\dagger}]^{R_1},
\end{equation}
where, $T_{2}$ is one of two partial transposition operations: $\langle i \alpha|U|j \beta \rangle=\langle j \alpha |U^{T_{1}}|i \beta \rangle=\langle i \beta |U^{T_{2}}|j \alpha \rangle$. The two-point correlation ``cross-correlation" is determined by the complementary channel as
\begin{equation}
    D_Q^{(12)} = \text{Tr}\left[\left( \mathbbm{1} \otimes \hat{b}_0\right) U^{\dagger} \left(\hat{a} \otimes \frac{\mathbbm{1}}{N} \right)U \right] = \text{Tr}_{2}(\hat{a}' \hat{b}_0)= \bra{\hat{a}^{\dagger}}\ket{\hat{b}_0}= \bra{\hat{a}^{\dagger}}\mathcal{M}_C(U)\ket{\hat{b}_0}.
\end{equation}
For dual unitary $U$ this does not decorrelate. It does for those $U$ which are such that their partial transpose is unitary, being referred to in some literature as ``T-duality" \cite{Bernoulli_Suhail}. Thus if $U$ is T-dual, the complementary channel $\mathcal{M}_C(U)$ is completely depolarizing. It is possible that there are unitary $U$ such that it is dual unitary as well as T-dual. Such matrices have been called 2-unitary \cite{Goyeneche2015} and in some contexts as ``perfect tensors" \cite{Hosur2016,Pastawski2015}. For $N=2$ it is known that there are no 2-unitary matrices, while they do exist for all $N>2$ \cite{Suhail_EnsembleDU,Goyeneche2015}. Channels corresponding to 2-unitary matrices are such that both the direct and complementary channels are completely depolarizing. In the context of brick-wall circuits they have been called Bernoulli \cite{frigg_ergodic} if built out of such unitaries \cite{Bernoulli_Suhail}.

Various diagrammatic methods such as tensor networks are useful for many-body physics. In the context of unitary circuits, especially dual-unitary ones they have resulted in great simplifications of calculations \cite{Dual_Prosen}. Although we do not study circuits in this paper, it is still useful perhaps to introduce them to motivate the classical correspondence as well. A bipartite unitary, its Hermitian conjugate, and  Eq.~(\ref{eq:opt_evolve}) can be represented as  
\begin{equation}
\br i \alpha |U|j \beta \kt = 
\begin{tikzpicture}[baseline=(current  bounding box.center),scale = 0.5]
\draw [thick,fill=blue,rounded corners=0.8pt] (1,1) rectangle (2,2); 
\node at (0.3,0.3) {$j$};
\node at (2.8,0.3) {$\beta$};
\node at (0.3,2.8) {$i$};
\node at (2.8,2.8) {$\alpha$};
\draw [thick] (2,2) -- (2.5,2.5);
\draw [thick] (1,2) -- (0.5,2.5);
\draw [thick] (1,1) -- (0.5,0.5);
\draw [thick] (2,1) -- (2.5,0.5);
\draw[thick] (1.5,1.8) -- (1.8,1.8) -- (1.8,1.5);
\end{tikzpicture},  \qquad  U^{\dagger} = 
\begin{tikzpicture}[baseline=(current  bounding box.center),scale = 0.5]
\draw [thick,fill=red,rounded corners=0.8pt] (1,1) rectangle (2,2); 
\draw [thick] (2,2) -- (2.5,2.5);
\draw [thick] (1,2) -- (0.5,2.5);
\draw [thick] (1,1) -- (0.5,0.5);
\draw [thick] (2,1) -- (2.5,0.5);
\draw[thick] (1.5,1.8) -- (1.8,1.8) -- (1.8,1.5);
\end{tikzpicture}, \qquad
    \begin{tikzpicture}[baseline=(current  bounding box.center),scale = 0.4]
\draw [thick,fill=blue,rounded corners=0.8pt] (1,1) rectangle (2,2); 
\draw [thick] (1,2) to[out=110, in=-110] (1,3);
\draw [thick] (2,2) to[out=70, in=-70] (2,3);
\draw[thick] (1.5,1.8) -- (1.8,1.8) -- (1.8,1.5);
\draw [thick,fill=red,rounded corners=0.8pt] (1,3) rectangle (2,4);
\draw [thick] (1,4) -- (0.5,4.5);
\draw [thick] (2,4) -- (2.5,4.5);
\draw[thick] (1.5,3.8) -- (1.8,3.8) -- (1.8,3.5);
\draw [thick] (1,1) -- (0.5,0.5);
\draw [thick] (2,1) -- (2.5,0.5);
\node at (0.4,2.5) {$\hat{a}$};
%\node at (2.5,2.5) {$\hat{\rho}$};
%\draw[thick, fill=black] (2.1,2.5) circle (0.1cm); 
\draw[thick, fill=black] (0.9,2.5) circle (0.1cm); 
\draw [thick] (2.5,0.5) to[out=70, in=-70] (2.5,4.5);
\end{tikzpicture}
=  
\begin{tikzpicture}[baseline=(current  bounding box.center),scale = 0.4] 

\draw [thick] (3.5,1) -- (3.5,4);
%\draw [thick] (4.5,1) -- (4.5,4);
\draw [thick] (3.5,1) -- (3.0,0.5);
\draw [thick] (3.5,4) -- (3,4.5);
%\draw [thick] (4.5,4) -- (5,4.5);
%\draw [thick] (4.5,1) -- (5,0.5);
\draw[thick, fill=black] (3.48,2.5) circle (0.1cm); 
\node at (3.0,2.7) {$\hat{a}'$};
\end{tikzpicture}, 
\end{equation}
where connecting legs is equivalent to contracting the indices.
To represent the channel in Eq.~(\ref{eq:qchan_act}), it is useful to use the doubled or folded operator representations \cite{Folded_Kos}, which is algebraically 
\begin{equation}
\mathcal{M}(U)=\bra{\fullmoon_{22'}} (U^{\dagger}_{12} \otimes U^T_{1'2'}) \ket{\fullmoon_{22'}}
=
\begin{tikzpicture}[baseline=(current  bounding  box.center), scale=1]
\def\eps{0.5};
\draw[very thick] (-4.25,0.5) -- (-3.25,-0.5);
\draw[very thick] (-4.25,-0.5) -- (-3.25,0.5);
\draw[ thick, fill=teal, rounded corners=0.8pt] (-4,0.25) rectangle (-3.5,-0.25);
\draw[thick, fill=white] (-3.25,-0.5) circle (0.075cm); 
%\node at (-3.0,-0.5) {$\hat{\rho}$};
\draw[thick, fill=white] (-3.25, 0.5) circle (0.075cm); 
\draw[thick] (-3.75,0.15) -- (-3.6,0.15) -- (-3.6,0);
\end{tikzpicture}
\end{equation}
Here $\begin{tikzpicture}
    \draw[thick] (0,0) -- (0.5,0); \draw[thick, fill=white] (0.5,0) circle (0.07cm);    \end{tikzpicture} = |\fullmoon\kt  $,  and 
\begin{equation}
\label{eq:doublegate}
\begin{tikzpicture}[baseline=(current  bounding  box.center), scale=1]
\def\eps{0.5};
\Wgateteal{-3.75}{0};
\Text[x=-3.75,y=-0.6]{}
\end{tikzpicture}
=
\begin{tikzpicture}[baseline=(current  bounding  box.center), scale=1]
\draw[thick] (-1.65,0.65) -- (-0.65,-0.35);
\draw[thick] (-1.65,-0.35) -- (-0.65,0.65);
\draw[ thick, fill=blue, rounded corners=0.8pt] (-1.4,0.4) rectangle (-.9,-0.1);
\draw[thick] (-1.15,0) -- (-1,0) -- (-1,0.15);
\draw[thick] (-2.25,0.5) -- (-1.25,-0.5);
\draw[thick] (-2.25,-0.5) -- (-1.25,0.5);
\draw[ thick, fill=red, rounded corners=0.8pt] (-2,0.25) rectangle (-1.5,-0.25);
\draw[thick] (-1.75,0.15) -- (-1.6,0.15) -- (-1.6,0);
\Text[x=-2.25,y=-0.6]{}
\end{tikzpicture}
= U_{12}^\dag\otimes U_{1'2'}^{T}\,\, ,
\end{equation}
Thus another characterization of dual unitarity is that iff $U_{12}$ is dual unitary $\bra{\fullmoon_{22'}} (U^{\dagger}_{12} \otimes U^T_{1'2'}) \ket{\fullmoon_{22'}}
= \ket{\fullmoon_{11'}}\bra{\fullmoon_{11'}}$.

States in the operator Hilbert space are represented by a black-filled circle,
$\begin{tikzpicture}
    \draw[thick] (0,0) -- (0.5,0); \draw[thick, fill=black] (0.5,0) circle (0.07cm);    \end{tikzpicture} =  |\hat{a} \kt $
In the folded picture, the diagrammatic representation for the evolution of the states, as in Eq.~(\ref{eq:opt_evolve}) is
\begin{equation}
\begin{tikzpicture}[baseline=(current  bounding  box.center), scale=1]
\def\eps{0.5};
\draw[very thick] (-4.25,0.5) -- (-3.25,-0.5);
\draw[very thick] (-4.25,-0.5) -- (-3.25,0.5);
\draw[ thick, fill=teal, rounded corners=0.8pt] (-4,0.25) rectangle (-3.5,-0.25);
\draw[thick] (-3.75,0.15) -- (-3.6,0.15) -- (-3.6,0);
\Text[x=-2.75,y=0.0, anchor = center]{$=$}
\draw[thick, fill=white] (-3.25,-0.5) circle (0.075cm); 
%\node at (-3.0,-0.5) {$\hat{\rho}$};
\draw[thick, fill=white] (-3.25, 0.5) circle (0.075cm);  
\Text[x=-1,y=0.0, anchor = center].
\draw[thick, fill=black] (-4.25,-0.5) circle (0.05cm); 
\node at (-4.5,-0.5) {$\hat{a}$};
\draw [very thick] (-2.1,0.5) to [out=60, in=-60] (-2.1,-0.5);
\node at (-2.1,-0.8) {$\hat{a}'$};
\draw[thick, fill=black] (-2.07,-0.5) circle (0.06cm); 
\end{tikzpicture}
\end{equation}
To distinguish it from an operator, folded gates are denoted with thick lines.

The complementary channel is 
\begin{equation}
\label{eq:diag_comp_qchan}
\mathcal{M}_{C}(U) = 
\begin{tikzpicture}[baseline=(current  bounding  box.center), scale=1]
\def\eps{0.5};
\draw[very thick] (-4.25,0.5) -- (-3.25,-0.5);
\draw[very thick] (-4.25,-0.5) -- (-3.25,0.5);
\draw[ thick, fill=teal, rounded corners=0.8pt] (-4,0.25) rectangle (-3.5,-0.25);
\draw[thick, fill=white] (-3.25,-0.5) circle (0.075cm); 
%\node at (-3.0,-0.5) {$\hat{\rho}$};
\draw[thick, fill=white] (-4.25, 0.5) circle (0.075cm); 
\draw[thick] (-3.75,0.15) -- (-3.6,0.15) -- (-3.6,0);
\end{tikzpicture}.
\end{equation}
Also, $\mathcal{M}_{C}(U) = \mathcal{M}(US)$, where $S$ is the swap operator $ S |i\kt |\alpha\kt = |\alpha\kt |i\kt$. If $U$ is T-dual (dual) then $US$ is dual (T-dual), hence the complementary channel of a T-dual (dual) operator is a channel for the corresponding dual (T-dual) unitary. There are two other channels $\mathcal{M}(SU)$ and $\mathcal{M}(SUS)$, and together these four channels determine the four possible correlation functions of single particle observables in a bipartite setting. We note that the definition of a channel and its complementary may be interchanged in some other papers, and notations such as $\mathcal{M}_{\pm}$ are also used \cite{Dual_Prosen,Bernoulli_Suhail}.

The diagrammatic representation of the correlation $D^{(11)}_Q$ is 
\begin{equation}
D^{(11)}_Q =\frac{1}{N}
\begin{tikzpicture}[baseline=(current  bounding box.center),scale = 0.4]
\draw [thick,fill=blue,rounded corners=0.8pt] (1,1) rectangle (2,2); 
\draw [thick] (1,2) to[out=110, in=-110] (1,3);
\draw [thick] (2,2) to[out=70, in=-70] (2,3);
\draw[thick] (1.5,1.8) -- (1.8,1.8) -- (1.8,1.5);

\draw [thick,fill=red,rounded corners=0.8pt] (1,3) rectangle (2,4);
\draw [thick] (1,4) -- (0.5,4.5);
\draw [thick] (2,4) -- (2.5,4.5);
\draw[thick] (1.5,3.8) -- (1.8,3.8) -- (1.8,3.5);
\draw [thick] (1,1) -- (0.5,0.5);
\draw [thick] (2,1) -- (2.5,0.5);
\node at (0.4,2.5) {$\hat{a}$};
%\node at (2.5,2.5) {$\hat{\rho}$};
\node at (1.0,4.7) {$\hat{a}_0$};
\draw[thick, fill=black] (0.9,2.5) circle (0.1cm); 
%\draw[thick, fill=black] (2.1,2.5) circle (0.1cm);
\draw[thick, fill=black] (0.75,4.25) circle (0.09cm);
\draw [thick] (2.5,0.5) to[out=70, in=-70] (2.5,4.5);
\draw [thick] (0.5,0.5) to[out=110, in=-110] (0.5,4.5);
\end{tikzpicture}\,\,\, =
\begin{tikzpicture}[baseline=(current  bounding  box.center), scale=1]
\def\eps{0.5};
\draw[very thick] (-4.25,0.5) -- (-3.25,-0.5);
\draw[very thick] (-4.25,-0.5) -- (-3.25,0.5);
\draw[ thick, fill=teal, rounded corners=0.8pt] (-4,0.25) rectangle (-3.5,-0.25);
\draw[thick, fill=white] (-3.25,-0.5) circle (0.075cm); 
\draw[thick, fill=black] (-4.25,-0.5) circle (0.065cm); 
\node at (-4.5,-0.5) {$\hat{a}$};
%\node at (-3.0,-0.5) {$\hat{\rho}$};
\draw[thick, fill=white] (-3.25, 0.5) circle (0.075cm); 
\node at (-4.5,0.5) {$\hat{a}_0$};
\draw[thick, fill=black] (-4.25,0.5) circle (0.065cm);
\draw[thick] (-3.75,0.15) -- (-3.6,0.15) -- (-3.6,0);
\end{tikzpicture}.
\end{equation}

The channels above have well-known properties \cite{Nielsen_Chuang_2010,Watrous_2018,Wilde_book_QI,davies1976quantum,Preskill1998}, when thought as trace preserving channels of quantum states. They are  complete positive, they map positive operators (say $\rho \in \mathcal{B}(\mathcal{H}_1^N)$) to positive operators. Completeness requires that $\rho$ can be entangled with a larger system on which the evolution acts trivially as an identity (hence not the subsystem $2$ considered here). Mathematically, complete positivity of $\mathcal{M}(U)$ one requires the realignment $\mathcal{M}(U)^{R_{2}}$ (or equivalently the realigment $R_1$) to be positive \cite{ECG_CP, Bengtsson_Zyczkowski_2006}, and this is evident from the forms in Eq.~(\ref{eq:qchan_act}) and Eq.~(\ref{eq:comp_qchan}). They are also {\em unital}, that is they map the maximally mixed state $\mathbbm{1}_{N}/N$ to itself. As the identity vectorizes in the standard maximally entangled state, and diagrammatically, the unitality condition is
\begin{equation}
\begin{tikzpicture}[baseline=(current  bounding  box.center), scale=1]
\def\eps{0.5};
\draw[very thick] (-4.25,0.5) -- (-3.25,-0.5);
\draw[very thick] (-4.25,-0.5) -- (-3.25,0.5);
\draw[ thick, fill=teal, rounded corners=0.8pt] (-4,0.25) rectangle (-3.5,-0.25);
\draw[thick] (-3.75,0.15) -- (-3.6,0.15) -- (-3.6,0);
%\draw[thick] (-3.75,0.15) -- (-3.6,0.15) -- (-3.6,0);
\Text[x=-2.75,y=0.0, anchor = center]{$=$}
\draw[thick, fill=white] (-4.25,-0.5) circle (0.075cm); 
\draw[thick, fill=white] (-3.25,-0.5) circle (0.075cm); 
\draw[very thick] (-2.25,0.5) -- (-2.25,-0.5);
\draw[very thick] (-1.25,-0.5) -- (-1.25,0.5);
\draw[thick, fill=white] (-2.25,-0.5) circle (0.075cm); 
\draw[very thick, fill=white] (-1.25,-0.5) circle (0.075cm); 
\end{tikzpicture},
\begin{tikzpicture}[baseline=(current  bounding  box.center), scale=1]
\def\eps{0.5};
\draw[very thick] (-4.25,0.5) -- (-3.25,-0.5);
\draw[very thick] (-4.25,-0.5) -- (-3.25,0.5);
\draw[ thick, fill=teal, rounded corners=0.8pt] (-4,0.25) rectangle (-3.5,-0.25);
\draw[thick] (-3.75,0.15) -- (-3.6,0.15) -- (-3.6,0);
\Text[x=-2.75,y=0.0, anchor = center]{$=$}
\draw[thick, fill=white] (-4.25,0.5) circle (0.075cm); 
\draw[thick, fill=white] (-3.25,0.5) circle (0.075cm); 
\draw[very thick] (-2.25,0.5) -- (-2.25,-0.5);
\draw[very thick] (-1.25,-0.5) -- (-1.25,0.5);
\draw[thick, fill=white] (-2.25,0.5) circle (0.075cm); 
\draw[thick, fill=white] (-1.25,0.5) circle (0.075cm); 
\end{tikzpicture},
\label{eq:unitality_diag}
\end{equation}

\subsection{\label{subsec:class_channel} Classical Channels from the Koopman operator}

We wish to study cases when $U$ has a semiclassical or classical limit and hence we might as well study the purely classical channels, which does not seem to be standard. A classical bipartite system, in its simplest setting, is separated into two degrees of freedom, and the evolution of functions on this four-dimensional phase space, $\Omega$, is given by a Koopman operator, $\hat{\mathcal{K}}$. The time subscript is now dropped as we will consider a discrete time dynamics whose period is taken as the unit of time. Consider a canonical transformation, $\Phi: (q_{1},p_{1},q_{2},p_{2}) \mapsto (q_{1}',p_{1}',q_{2}',p_{2}')$. The matrix elements of the Koopman operator for such a transformation are  
\begin{equation}
    \label{eq:koop_define}
    \begin{split}
    \br \xi_{1} \xi_{2}|\hat{\mathcal{K}}|\eta_{1} \eta_{2}\kt & = \delta [\eta_{1}-\Phi_{1}(\xi_{1},\xi_{2})]\,\delta [\eta_{2}-\Phi_{2}(\xi_{1},\xi_{2})]\\ & \equiv K(\eta_1,\eta_2,\xi_1,\xi_2),
    \end{split}
\end{equation}
where, $\xi_i$ and $\eta_i$ represent the phase space coordinates of each degree of freedom $(q_i,p_i)$. The action of $\hat{\mathcal{K}}$ on a function $f$ is given by the composition $f \circ \Phi$, and is found as 
   $\hat{\mathcal{K}} f(\xi_{1},\xi_2) = \bra{\xi_1 
   \xi_2} \hat{\mathcal{K}} \ket{ f}= f(\xi_{1}',\xi_{2}')$. The Koopman $\hK$ is a classical unitary channel.
   
Just as in the quantum case, where we defined four, in general, non-unitary channels we can define four classical channels from the Koopman $\hK$, or its kernel function $K$ as:
\beq
\begin{split}
\bra{\xi_1}\hK_{R_1}\ket{\eta_1}=&
\int K\, d\xi_2 d\eta_2, \; \bra{\xi_2}\hK_{R_2}\ket{\eta_2}=
\int K\, d\xi_1 d\eta_1, \\
\bra{\xi_2}\hK_{\Gamma}^{+}\ket{\eta_1}=&
\int K\,d\xi_1 d\eta_2,\, \bra{\xi_1}\hK_{\Gamma}^{-}\ket{\eta_2}=
\int K\,d\xi_2 d\eta_1
\end{split}
\label{eq:koop_channel}
%\label{eq:Maps}
\eeq 
where the integrations range is over all the relevant spaces. 
From the specific form of the Koopman in Eq.~(\ref{eq:koop_define}) we find more explicitly that 
\beq
\begin{split}
\bra{\xi_1}\hK_{R_1}\ket{\eta_1}=&
\int \delta [\eta_{1}-\Phi_{1}(\xi_{1},\xi_{2})] \, d \xi_2, \; \bra{\xi_2}\hK_{R_2}\ket{\eta_2}=
\int \delta [\eta_{2}-\Phi_{2}(\xi_{1},\xi_{2})] \, d \xi_1, \\
\bra{\xi_2}\hK_{\Gamma}^{+}\ket{\eta_1}=&
\int \delta [\eta_{1}-\Phi_{1}(\xi_{1},\xi_{2})]\, d \xi_1,\, \bra{\xi_1}\hK_{\Gamma}^{-}\ket{\eta_2}=
\int \delta [\eta_{2}-\Phi_{2}(\xi_{1},\xi_{2})] \, d \xi_2.
\end{split}
\label{eq:koop_channel2}
%\label{eq:Maps}
\eeq

These channels can be used to find various correlation functions of observables localized on single particles, just as in the quantum case.  
For example, the dynamical correlation between a time evolved single particle observable, say given by $\hat{\mathcal{K}}f(\xi_1)$ and one that is not, say $g(\xi_1)$, is 
\begin{equation}
    \label{classical_corr}
    D_{C}^{(11)}= \bra{g}\bra{F_u} \hat{\mathcal{K}} \ket{f}\ket{F_u}=\int_{\Omega_1} \br g|\xi_{1} \kt \br \xi_{1}|\hat{\mathcal{K}}_{R_1}|\eta_{1}\kt \br \eta_{1}|f \kt \; d\xi_{1}d\eta_{1}=\bra{g}\hat{\mathcal{K}}_{R_1} \ket{f}, 
\end{equation}
where $\bra{\xi_2}\ket{F_u}=F_{u}(\xi_{2})=1$ for all $\xi_2$ is the uniform function on the phase space of the second degree of freedom. It's diagrammatic representation  is 
\begin{equation}
\label{eq:diag_koop_chan}
\hat{\mathcal{K}}_{R_1} = 
\begin{tikzpicture}[baseline=(current  bounding  box.center), scale=1]
\def\eps{0.5};
\draw[thick] (-4.25,0.5) -- (-3.25,-0.5);
\draw[ thick] (-4.25,-0.5) -- (-3.25,0.5);
\draw[ thick, fill=myblue, rounded corners=0.8pt] (-4,0.25) rectangle (-3.5,-0.25);
\draw[thick, fill=white](-3.25,-0.5) circle (0.075cm); 
%\node at (-3.0,-0.5) {${\rho}$};
\draw[thick, fill=white] (-3.25, 0.5) circle (0.075cm);
\draw[thick] (-3.75,0.15) -- (-3.6,0.15) -- (-3.6,0);
\end{tikzpicture}
\end{equation}
where the square represents the Koopman operator, and $ \begin{tikzpicture}
    \draw[thick] (0,0) -- (0.5,0); \draw[thick, fill=white] (0.5,0) circle (0.07cm);    
\end{tikzpicture} = F_{u}(\xi).$
Comparing the expression of $D_{C}^{(11)}$ with $D_{Q}^{(11)}$, in Eq.~(\ref{eq:corr_fold}) we see that $\hat{\mathcal{K}}_{R_1}$ is the classical channel associated with the Koopman unitary channel, $\hat{\mathcal{K}}$, just as the quantum channel $\mathcal{M}(U)$
is associated with the unitary channel $U^{\dagger} \otimes U^T$. Both the operators are in general sub-unitary and while their diagrammatic representations are identical, they hide significant differences. We also note that the ``folded" representation is natural to classical mechanics as there is no clear equivalent of $U$ itself.
The channel complementary to $\hat{\mathcal{K}}_{R_1}$ is 
\begin{equation}
    \label{eq:comp_koop_channel}
    \hat{\mathcal{K}}_{\Gamma}^+  = 
\begin{tikzpicture}[baseline=(current  bounding  box.center), scale=1]
\def\eps{0.5};
\draw[thick] (-4.25,0.5) -- (-3.25,-0.5);
\draw[ thick] (-4.25,-0.5) -- (-3.25,0.5);
\draw[ thick, fill=myblue, rounded corners=0.8pt] (-4,0.25) rectangle (-3.5,-0.25);
\draw[thick, fill=white](-3.25,-0.5) circle (0.075cm); 
%\node at (-3.0,-0.5) {${\rho}$};
\draw[thick, fill=white] (-4.25, 0.5) circle (0.075cm); 
\draw[thick] (-3.75,0.15) -- (-3.6,0.15) -- (-3.6,0);
\end{tikzpicture}.
\end{equation}

The Unitality of the quantum channels were represented in Eq.~(\ref{eq:unitality_diag}) and the exact same diagrams should hold for the classical Koopman operator. In fact rather than the unitarity, the unital condition is important in the context of many-body systems for example \cite{Folded_Kos}.
A classical channel is unital if it leaves the uniform function invariant, and in this sense the condition of the first diagram from Eq.~(\ref{eq:unitality_diag}) is actually trivial for any transformation, not necessarily canonical, as it follows from Eq.~(\ref{eq:koop_define}) that 
\beq
\bra{\xi_1\xi_2}\hK \ket{F_u(\eta_1) F_u(\eta_2)}=\int_{\Omega_2} \br \xi_{1} \xi_{2}|\hat{\mathcal{K}}|\eta_{1} \eta_{2}\kt \, d \eta_1 d \eta_2 =1.
\eeq
The other unitality condition, the second of  Eq.~(\ref{eq:unitality_diag}) follows from the fact that the transformation $\Phi$ is canoncial, invertible and preserves the Lebesgue measure (Liouville theorem of classical mechanics).

If in addition to the unitality, $\hK$ satisfies:
\beq
\begin{tikzpicture}[baseline=(current  bounding  box.center), scale=1]
\def\eps{0.5};
\draw[thick] (-4.25,0.5) -- (-3.25,-0.5);
\draw[thick] (-4.25,-0.5) -- (-3.25,0.5);
\draw[ thick, fill=myblue, rounded corners=0.8pt] (-4,0.25) rectangle (-3.5,-0.25);
\draw[thick] (-3.75,0.15) -- (-3.6,0.15) -- (-3.6,0);
\Text[x=-2.75,y=0.0, anchor = center]{$=$}
\draw[thick, fill=white] (-3.25,-0.5) circle (0.075cm); 
\draw[thick, fill=white] (-3.25, 0.5) circle (0.075cm); 
\draw[thick] (-1.25,0.5) -- (-2.25, 0.5);
\draw[thick] (-1.25,-0.5) -- (-2.25,-0.5);
\draw[thick, fill=white] (-1.25, 0.5) circle (0.075cm); 
\draw[thick, fill=white] (-1.25,-0.5) circle (0.075cm); 
\Text[x=-1,y=0.0, anchor = center]{,}
\end{tikzpicture}
\qquad\,\, 
\begin{tikzpicture}[baseline=(current  bounding  box.center), scale=1]
\def\eps{0.5};
\draw[thick] (-4.25,0.5) -- (-3.25,-0.5);
\draw[thick] (-4.25,-0.5) -- (-3.25,0.5);
\draw[ thick, fill=myblue, rounded corners=2pt] (-4,0.25) rectangle (-3.5,-0.25);
\draw[thick] (-3.75,0.15) -- (-3.6,0.15) -- (-3.6,0);
\Text[x=-2.75,y=0.0, anchor = center]{$=$}
\draw[thick, fill=white] (-4.25,0.5) circle (0.075cm); 
\draw[thick, fill=white] (-4.25,-0.5) circle (0.075cm); 
\draw[thick] (-2.25,0.5) -- (-1.25,0.5);
\draw[thick] (-2.25,-0.5) -- (-1.25,-0.5);
\draw[thick, fill=white] (-2.25,-0.5) circle (0.075cm); 
\draw[thick, fill=white] (-2.25,0.5) circle (0.075cm);
\end{tikzpicture}.
\label{eq:dualunitality_diag}
\eeq
it has been referred to as being a dual-Koopman operator 
\cite{Dual_Koopman_Arul}. It is T-dual if 
\beq
\begin{tikzpicture}[baseline=(current  bounding  box.center), scale=1]
\def\eps{0.5};
\draw[thick] (-4.25,0.5) -- (-3.25,-0.5);
\draw[thick] (-4.25,-0.5) -- (-3.25,0.5);
\draw[ thick, fill=myblue, rounded corners=0.8pt] (-4,0.25) rectangle (-3.5,-0.25);
\draw[thick] (-3.75,0.15) -- (-3.6,0.15) -- (-3.6,0);
\Text[x=-2.75,y=0.0, anchor = center]{$=$}
\draw[thick, fill=white] (-3.25,-0.5) circle (0.075cm); 
\draw[thick, fill=white] (-4.25, 0.5) circle (0.075cm); 
\draw[thick] (-1.25,0.5) -- (-2.25, 0.5);
\draw[thick] (-1.25,-0.5) -- (-2.25,-0.5);
\draw[thick, fill=white] (-2.25, 0.5) circle (0.075cm); 
\draw[thick, fill=white] (-1.25,-0.5) circle (0.075cm); 
\Text[x=-1,y=0.0, anchor = center]{,}
\end{tikzpicture}
\qquad\,\, 
\begin{tikzpicture}[baseline=(current  bounding  box.center), scale=1]
\def\eps{0.5};
\draw[thick] (-4.25,0.5) -- (-3.25,-0.5);
\draw[thick] (-4.25,-0.5) -- (-3.25,0.5);
\draw[ thick, fill=myblue, rounded corners=0.8pt] (-4,0.25) rectangle (-3.5,-0.25);
\draw[thick] (-3.75,0.15) -- (-3.6,0.15) -- (-3.6,0);
\Text[x=-2.75,y=0.0, anchor = center]{$=$}
\draw[thick, fill=white] (-3.25,0.5) circle (0.075cm); 
\draw[thick, fill=white] (-4.25,-0.5) circle (0.075cm); 
\draw[thick] (-2.25,0.5) -- (-1.25,0.5);
\draw[thick] (-2.25,-0.5) -- (-1.25,-0.5);
\draw[thick, fill=white] (-1.25,0.5) circle (0.075cm); 
\draw[thick, fill=white] (-2.25,-0.5) circle (0.075cm);
\end{tikzpicture}.
\label{eq:Gammadualunitality_diag}
\eeq
 If $\hat{K}$ is dual, then $\hat{K}_{R_i}=\ket{F_u}\bra{F_u}$  and if it is $\Gamma$-dual, then $\hat{K}_{\Gamma}^{\pm}=\ket{F_u}\bra{F_u}$, that is they are rank-1 projectors onto the uniform function. In general they are sub-unitary operators, but in any case they have $F_u(\xi)$ as a ``trivial" eigenfunction with eigenvalue $1$. If all of the above, spatial and temporal unitality conditions are met, the operator $\hK$ is a ``perfect Koopman operator", after perfect tensors that have similar properties in quantum information theory \cite{Pastawski2015,Bernoulli_Suhail,Hosur2016}.

\subsection{An example}

Consider a kicked two-degree of freedom system with the Hamiltonian:
\beq
\label{eq:Hamil}
H=\frac{1}{2}(p_1^2+p_2^2)+ V(q_1,q_2)\sum_{n=-\infty}^{\infty} \delta(t-n).
\eeq
 The canonical transformation relating the phase space variables just before the two kicks is the 4-dimensional symplectic map $T$:
\begin{equation} 
\label{eq:gen_map}
\begin{split}
    & q_{i}' = q_{i}  + p_{i}'
    \\
    & p_{i}' = p_{i} - \frac{\partial V}{\partial q_{i}}(q_{1},q_{2})
\end{split}
\end{equation}
where, $i=\{1,2\}$. 
The potential, $V(q_{1},q_{2})=V_{1}(q_1)+V_{2}(q_2)+V_{12}(q_1,q_2)$ where, $V_{i}$ is the local potential on the subsystem $i$ and $V_{12}$ gives the interaction potential between the subsystems $1$ and $2$. We consider the case when $q_i$ are angle variable and restrict the motion of the particle on a two torus. The symmetry along the momentum direction implies that translations in momentum by the periodicity of $q_i$ commutes with the dynamics.
Hence, the modulo operation (we scale so that it is mod 1) can be implied in all the four equations in Eq.~(\ref{eq:gen_map}). 
This makes the classical phase-space the compact 4-torus and the quantum mechanics is finite dimensional. The classical Koopman operator $\hat{\mathcal{K}}$, and the Floquet unitary operator $U$ quantizing this canonical transformation are both T-dual. Thus the time-discretization of natural two-degrees of freedom Hamiltonians are of this special kind.

The map in Eq.~(\ref{eq:gen_map}) can be written as a composition of three canonical maps $T=  (T_1 \circ T_2) \circ T_{12}$, where 
$T_i: (q_i'=q_i+p_i', \, p_i'=p_i- dV_i(q_i)/dq_i)$ are ``local maps" and $T_{12}: (q_i'=q_i, \, p_i'=p_i -\partial V_{12}(q_1,q_2)/\partial q_i)$ is the pure interaction impulse.
This implies that $\hK=(\hK_1 \otimes \hK_2)\hK_{12}$ and $U=(U_1\otimes U_2)U_{12}$. The properties of duality (or T-duality) are nonlocal properties: they are not affected by local canonical or unitary transformations \cite{Dual_Koopman_Arul}. Therefore it is sufficient to prove that $\hK_{12}$ and $U_{12}$ are T-dual.

If we denote $\xi_i=(q_i,p_i)$, $\eta_i=(q_i',p_i')$, then $\Phi_i:(q_i, p_i-\partial V_{12}(q_1,q_2)/\partial q_i)$, from Eq.~(\ref{eq:koop_channel2}) we find that
\beq
\label{eq:KGdual}
\begin{split}
\bra{\xi_2} \hK_{12 \, \Gamma}^{+}\ket{\eta_1}
& =\int \delta[\eta_{1}-\Phi_{1}(\xi_{1},\xi_{2})]\,d \xi_1\\
&=\int \delta\left(q_1'-q_1\right)\, \delta\left[ p_1'-p_1+\partial V_{12}(q_1,q_2)/\partial q_1 \right] dq_1 dp_1=1.
\end{split}
\eeq
Hence the channel $\hK_{12}$ is T-dual, and so is $\hK$.
Similarly, the quantum interaction unitary, $U_{12}=\exp(-i V_{12}(q_1,q_2)/\hbar)$ is T-dual. Thus it is diagonal in the position basis: $\br i\alpha  | U_{12} | j\beta \kt = e^{i\phi_{j\beta}} \delta_{\alpha,\beta}\delta_{i,j}$. Therefore, $\br i\alpha  | U_{12}^{T_{2}} | j\beta \kt = e^{i\phi_{j\alpha}} \delta_{\alpha,\beta}\delta_{i,j}$, and hence $U_{12}=U_{12}^{T_2}$, and $U_{12}$ is not only T-dual, its T-dual is itself. There is nothing special about the position representation, under any local basis change, such as to momentum, the operator is in general no longer diagonal but is locally connected to a diagonal. All such unitaries are T-dual, and hence $U$ is also T-dual \cite{Bernoulli_Suhail}. The nontrivial channels are $\hK_{R_1}$ and $\hK_{R_2}$, the other two are maximally mixing.

\subsection{\label{subsec:class_chan_noise} Classical Channels as stochastic ones}

Once the channels are defined, their powers and spectra are interesting. It is clear that the powers of $\hK$ and its spectra are relevant to the classical canonical transformation $\Phi$. It is not immediately clear what the role of the other four channels are. In fact they were defined in \cite{Dual_Koopman_Arul} as they decide {\em many-body} correlations of 1+1 brick-wall, just as for the quantum case \cite{Dual_Prosen}. If the elements are dual-Koopman (or dual-unitary in the quantum case), the correlations of single particles vanish everywhere except on the light cone, and  appropriate channels (the ``$\Gamma$" ones) then  determine the rate of decay on this \cite{Dual_Prosen}. Hence ergodic properties of many-body quantum and classical systems can be at least partially determined by these channels. However, we wish to go the other way and interpret these non-unitary channels as Koopman operators corresponding to a {\em noisy or random single-particle dynamics}.

This is quite evident from the form of the channels in Eq.~( \ref{eq:koop_channel2}). For example, the $\hK_{R_1}$ map corresponds to a one-particle Koopman whose underlying transform is $\Phi_1$ and the variable $\xi_2$, from the other degree of freedom is being intergrated (traced) out and can be interpreted as noise as far as the first degree of freedom is concerned. A very similar interpretation holds for $\hK_{R_2}$.
Considering the case of the example in Eq.~(\ref{eq:gen_map})  the canonical transformation only of a single particle (dropping the indices for now) is 
\begin{equation}
\begin{split}
    \label{eq:noisy_map}
    & q'=q + p'
    \\
    & p'=p - \frac{\partial V}{\partial q}(q,z)
\end{split}
\end{equation}
where $z$ is a parameter. The unitary Koopman operator corresponding to this transformation is $\br \xi|\hat{\mathcal{K}}_z|\eta\kt = \delta[\eta-\phi_{z}(\xi)]$. Notation has been changed: $\phi_z(\xi)\equiv \Phi_{1}(\xi,z)$ for later convenience, the vestige of this being a first degree of freedom is dropped, while the variable corresponding to the second degree, namely $z$, has been separated as a subscript. If $z$ is intepreted as a random varaible, the {\em stochastic Koopman operator} $\mathbbm{E}_z \hK_z$, where the expectation value is over some distribution (in this case uniform) of $z$ may be identified with $\hK_{R_1}$. Explicitly
\begin{equation}
    \label{eq:noisy_koop}
    \br \xi|\hat{\mathcal{K}}_{R_1}|\eta\kt = \mathbbm{E}_{z} \delta[\eta-\phi_{z}(\xi)]=\int  \delta[\eta-\phi_{z}(\xi)] \, dz,
\end{equation}
or in terms of action on functions $\hK_{R_1} f(\xi)=\mathbbm{E}_z \,f[\phi_z(\xi)]$.
Thus these channels have a semigroup property enjoyed by  stochastic Koopman operators \cite{Gaspard_1998,Mezic_koopbook,brunton2019data}. Simply,
\beq
\label{eq:koop_composition}
\hK_{R_1}^2f(\xi)=\br \xi|\hat{\mathcal{K}}_{R_1}^2|f \kt=\mathbbm{E}_{z',z}f[\phi_{z'} \circ \phi_z(\xi)]=\hK_{R_1} \left[\hK_{R_1} f(\xi)\right].
\eeq
Thus, the classical channel, $\hat{\mathcal{K}}_{R_1}$, can be interpreted as the Koopman operator corresponding to a noisy system with one degree of freedom. Its spectrum is in general inside the unit circle and desribes the decay of correlations with time wrought by the noise. 
The nature of the noise depends on the initial state of the second degree of freedom or ``environment", and $z$ being chosen from a uniform distribution is a consequence of this being a uniform function.

Note that the noise in the consideration above was origintaing from the second degree of freedom, which could be weakly coupled to the first, thus there is a control on the degree to which noise affects the system. This is also the case for the channel $\hK_{R_2}$, however the other two channels are qualitatively different.
The channels $\hK_{\Gamma}^{\pm}$ take a function from one subsystem to another, and the ``noise" originates from the parent degrees of freedom. For example:
\beq
\hK_{\Gamma}^+: f(\xi_1) \mapsto g(\xi_2)=\int f[ \Phi_1(\xi_1,\xi_2)] \, d \xi_1.
\eeq
As the map in Eq.~(\ref{eq:gen_map}) is T-dual, see Eq.~(\ref{eq:KGdual}), it follows that in this case
\beq
\hK_{\Gamma}^+ f(\xi_1)=\int f(\xi_1) d\xi_1,
\eeq
a constant function on the second degree of freedom. This is the classical equivalent of the completely depolarizing channel which is of rank-1. However, of course there are dual-unitary Koopman operators which are of rank-1 also, and this can be the case for $\hK_R$ channels too. However, these are trivially connected to T-dual ones via, unsurprisingly, the swap operation that exchanges the degrees of freedom. Formally if we define the swap Koopman operator to be $\hK_S F(\xi_1,\xi_2)=F(\xi_2,\xi_1)$ for all functions $F$, then the kernel corresponding to $\hK_S \hK$ is 
\begin{equation}
    \label{eq:koopSwap_define}
    \begin{split}
    \br \xi_{1} \xi_{2}|\hK_S \hat{\mathcal{K}}|\eta_{1} \eta_{2}\kt & = \delta [\eta_{1}-\Phi_{2}(\xi_{1},\xi_{2})]\,\delta [\eta_{2}-\Phi_{1}(\xi_{1},\xi_{2})]\\ & \equiv K_S(\eta_1,\eta_2,\xi_1,\xi_2),
    \end{split}
\end{equation}
The following holds then from Eq.~(\ref{eq:koop_channel}) and Eq.~(\ref{eq:koop_channel2}).
\beq
\hK_{R_1}'=\hK_{\Gamma}^-, \; \hK_{R_2}'=\hK_{\Gamma}^+,\; \hK_{\Gamma}^{+'}=\hK_{R_2},\; \hK_{\Gamma}^{-'}=\hK_{R_1},
\eeq
where the new channels are primed. It is clear therefore that T-dual channels will turn to being dual and vice-versa. Thus for these class of canonical transformations, the ``noise" is from the same source as the dynamics and we cannot expect there to be a ``control" in this case, and hence complete depolarizing channels occur.  

If we consider the canonical transform in Eq.~(\ref{eq:gen_map}) and follow it up with swap, it leads to dual Koopman operators and these have highly chaotic properties especially for many-body systems constructed from them. However, the role of the nonlocal swap operator is crucial. There are recent constructions of many-body models that are all swap operations, except just one ``impurity" that is T-dual \cite{Fritzsch2023}, and these show large chaos as well. This constitutes perhaps a different take on the spacetime dual gates that are currently very popular as toy-models of many-body quantum chaos \cite{Akila_2016,BorisGutkin2020,Peter2020,Dual_Prosen,Bertini_SFF2018,Piroli,Pieter_Lamacraft,Gopalkrishnan2019,Bernoulli_Suhail}.

\subsection{\label{subsec:quan_chann_noise} Quantum Channels and Noise}

The quantum Floquet operator corresponding to the family of Hamiltonians given in Eq.~(\ref{eq:Hamil}) is of the form
$U=(U_1\otimes U_2) U_{12}$, where $U_i$ are the local evolutions, including the single particle potentials $V_i(q_i)$, and $U_{12}$ is the interaction $\exp[-i V_{12}(q_1,q_2)/\hbar]$. The quantum channel, $\mathcal{M}(U)$ follows from Eq.~(\ref{eq:opt_evolve}) and is 
\begin{equation}
    \label{eq:gen_quan_channel}
    \mathcal{M}(U)=\frac{1}{N}[U_{12}^{R_{1}\dagger} U_{12}^{R_{1}}]^{R_{2}}(U_{1}^{\dagger} \otimes U_{1}^{\text{T}}) = \mathcal{M}(U_{12}) (U_{1}^{\dagger} \otimes U_{1}^{\text{T}}).
\end{equation}
The non-unitary part of the channel, $\mathcal{M}(U_{12})$, arises from the interaction part. As in the classical case, we may interpret this channel as arising due to ``noise" from the second degree of freedom. To understand this better, we focus on the non-unitary part separately. 

The interaction unitary, $U_{12}$ is diagonal in the position basis which we will use. Without specifying the explicit form of the phases for now we write: $\br k\alpha | U_{12} |l\beta \kt =\exp(i\phi_{k\beta}) \delta_{\alpha\beta}\delta_{kl}$.
Here the position basis used is a finite, $N$ dimensional one.
 The matrix elements of $\mathcal{M}(U_{12})$ are
\begin{equation}
    \label{eq:mat_noise_qchannel}
    \br k' k | \mathcal{M}(U_{12}) |l' l \kt = \frac{1}{N} \sum_{\alpha=0}^{N-1} \exp[i (\phi_{l\alpha}-\phi_{l'\alpha})] \delta_{kl}\delta_{k'l'}.
\end{equation}
The following Krauss representation \cite{Nielsen_Chuang_2010,Preskill1998} is immediate:
\beq
\label{eq:KraussM}
\mathcal{M}(U_{12})=\frac{1}{N} \sum_{\alpha=0}^{N-1} D_{\alpha}^{\dagger} \otimes D_{\alpha}^T= \sum_{\alpha=0}^{N-1} E_{\alpha} \otimes E_{\alpha}^*,
\eeq
where $ \bra{k} D_{\alpha} \ket{l}=\exp(i \phi_{k \alpha}) \delta_{kl}$, is a diagonal unitary of dimension $N$, and $E_{\alpha}=D^{\dagger}_{\alpha}/\sqrt{N}$ may be referred to as the ``noise operators" corresponding to the channel $\mathcal{M}(U_{12})$ and hence to $\mathcal{M}(U)$ itself. Note that the index $\alpha$ refers to the second degree of freedom and hence when we are summing over this, we are tracing it out, and the individual unitary operators $D_{\alpha}$ are now applied with equal probability. We note that this is exactly the case with the Koopman operators just discussed above.

As a technical aside, we note that the channel $\mathcal{M}$ is an example of what has been referred to in the literature as a ``Hadamard" or Schur product channel \cite{King2005_hadamard,Seung1995,Chi1997}, as its action on a state (or operator) is via the matrix element wise multiplication: if $\ket{\rho'}=\mathcal{M}(U_{12}) \ket{\rho}$, then $\rho'= A \circ \rho$, where $A$ is the positive matrix, whose diagonal entries are uniformly $1$:  
\begin{equation}
    \label{eq:corr_matrix}
    A_{k'k} = \frac{1}{N}  \sum_{\alpha} \exp[i (\phi_{k\alpha}-\phi_{k'\alpha})].
\end{equation}
It has been shown in \cite{Chi1997,Seung1995,Dephasing_superchannel_karol} that any channel, $\mathcal{E}_{D}$ is a dephasing channel if and only if it is a Schur-product channel. A quantum channel that leaves the density matrix in a fixed distinguished basis invariant is known as a dephasing channel: $ \br i |\mathcal{E}_{D}(\rho) | i \kt = \br i|\rho| i\kt$.
It is clear from the above that such a distinguished basis for $\mathcal{M}$ is the position basis.
For a dephasing channel, there is a loss of coherence, and such a noise model dominates in the case of superconducting qubits.  

\section{\label{sec:spectra}Spectral Properties of the Channels}

This section looks at the spectrum of the classical and quantum channels and quantum-classical correspondence. We recall for notation and convenience some key properties.
The spectral decomposition of the two quantum and classical channels we concentrate on are written as:  
\begin{equation}
        \label{eq:eigval_qchan}
    \mathcal{M}(U) =  \sum_{i =0}^{N^2-1} \lambda_{i}^{Q}|\mathcal{X}^{R}_{i}\kt \br \mathcal{X}^{L}_{i}|, \;
    \hat{\mathcal{K}}_{R_1}=\sum_{i =0}^{\infty} \lambda_{i}^{C}|\mathcal{\phi}^{R}_{i}\kt \br \mathcal{\phi}^{L}_{i}|.
    \end{equation}
Here $\br \mathcal{X}^{L}_{i}|$ and $ |\mathcal{X}^{R}_{i}\kt$ are the left and the right eigenmodes of the quantum channel,
We will also consider  the matrix forms of the modes, denoted $\mX_i^{R,L}$. The usual orthogoinality relations are replaced by  $ \br \mathcal{X}^{L}_{i} | \mathcal{X}^{R}_{j}\kt  = \tr(\mX_i^{\dagger} \mX_j)= \delta_{ij}$.  
In both the channels $|\lambda^{Q,C}_i|\leq 1$ and if the eigenvalues are complex, they come in complex conjugate pairs. For a unital channel, such as the above, the largest and ``trivial" eigenvalue is $\lambda_{0}^{Q,C} = 1$ and the eigenmodes, both left and right, corresponding to this eigenvalue is $|\mathbbm{1}\kt$.
Thus, all other eigenmodes of the quantum channel are traceless. The eigenmodes corresponding to the real eigenvalues are hermitian, and for complex eigenvalues, the operator $\mathcal{X}_{i}^{R \dagger} = \mathcal{X}_{i}^{L} $. 
Similarly for the classical channel, except that this is infinite dimensional,  $\br \phi^{L}_{i}|$ and $|\phi^{R}_{i}\kt$ are the left and the right eigenmodes of the classical channel $\hK_{R_1}$. In the classical case, although a discrete sum is written, it can have a much more complex spectrum. For practical purposes we truncate the infinite dimensional operator and ensure that the significant eigenvalues are converged and do not change with truncation size.

\subsection{\label{subsec:model} The coupled standard map and the channels}

We now turn to a concrete example to study the details of the spectra of the channels. A well-studied model is provided by the coupled standard map or kicked rotor, obtained by considering the potential in Eq.~(\ref{eq:Hamil}) to be
\beq
V(q_1,q_2)=-\frac{1}{4 \pi^2}\left[\alpha_1 \cos(2 \pi q_1)+\alpha_2 \cos(2 \pi q_2) + b \, \cos[ 2 \pi(q_1+q_2) ] \right]. 
\eeq
The resulting four-dimensional map from Eq.~(\ref{eq:gen_map}) is 
\begin{equation} 
\label{eq:standard_map}
\begin{split}
    & q_{i}' = q_{i}  + p_{i}'
    \\
    & p_{i}' = p_{i} -\frac{1}{2 \pi}[
    \alpha_i \sin(2 \pi q_i)+
    b \sin[2 \pi(q_1+q_2)]], 
\end{split}
\end{equation}
with $(i=1,2)$. It can be restricted with unit periodicity along all phase-space coordinates to act on $T \times T$, where $T$ is a unit two-torus. While the full coupled standard map on the the 4-torus is hard to visualize \cite{Ritcher_4d,Lange_4d}, the uncoupled standard maps with $b=0$ are textbook examples, with a well-known phenomenology which we now briefly recap. We use $\alpha$ to indicate the kicking strength against the traditional $K$ to avoid notational chaos. If $\alpha=0$ as well, the dynamics is integrable, being just free motion on circles, and the angular momentum $p$ is a constant of motion. If $\alpha \ll 1$, the dynamics is predominantly regular being dominated by KAM (Kolmogorov Arnold Moser) tori, spanning the angle space, with slowly varying momenta. However, the nonlinear resonances grow in size around periodic orbits and overlap to generate larger areas of chaotic motion. The phase space get increasingly mixed as KAM tori break, and around $\alpha=1$ the last of them breaks giving rise to diffusion in momentum. The stable fixed point and its resonance dominates the small $\alpha$ regime and becomes unstable only at $\alpha=4$, when the rest of the phase space is already quite chaotic. For $\alpha >5$ the dynamics is increasinly chaotic with the measure of regular regions tending to zero, and the Lyapunov exponent being $\approx \ln (\alpha/2)$. However, this is not monotonic as regular island form ``out of the blue". For example, there are so-called accelerator modes that are formed when $2 \pi l \leq \alpha \leq \sqrt{(2 \pi l)^2+16}$, where $l=1,2,\cdots$, the most prominent one happening for $l=1$. There are other occurences of tangencies of stable and unstable manifolds that can throw up small islands, for example while $\alpha=8$ and $\alpha=10$ can look completely chaotic, at a small interval around $\alpha=9.26$ regular islands show up \cite{FTSE_Arul}. All of these affect the quantum and classical spectra of the channels we are discussing.

In order to represent the explicit form of the classical channel, we use the Fourier basis   defined as, $\br q,p | F_{m,n}\kt \equiv \br q,p|m,n\kt = \exp[2\pi i (qm+np)]$, $m,n= 0,\pm 1,...$.  The Koopman operator for the transformation in Eq.~(\ref{eq:standard_map}) has the form , $\hat{\mathcal{K}}=(\hat{\mathcal{K}}_{1} \otimes \hat{\mathcal{K}}_{2}) \hat{\mathcal{K}}_{12}$, and the matrix elements in the Fourier basis are \cite{Dual_Koopman_Arul}
\begin{align}
&\br{m_i',n_i'}|\hat{\mathcal{K}}_i|{m_i,n_i}\kt=J_{m_i-m_i'}(\alpha_i n_i') \, \delta_{n_i',m_i+n_i}, \label{eq:Koopman1}\\
& \br{m_1',n_1';m_2',n_2'}|\hat{\mathcal{K}}_{12}|{m_1,n_1;m_2,n_2}\kt= 
\delta_{n_1,n_1'}\delta_{n_2,n_2'}\delta_{m_1-m_1',m_2-m_2'}\, J_{m_1-m_1'}[b(n_1+n_2)],
\label{eq:Koopman_SM}
\end{align}
where $J_{\nu}(x)$ is the Bessel function of the first kind. Noting that integration over phase space results in picking only the $(m=0,n=0)$ mode, the matrix elements of the classical channel $\hK_{R_1}$ in Eq.~(\ref{eq:koop_channel}) are
\begin{align}
\label{eq:class_chan_uni}
    & \br m_1', n_1' |\hat{\mathcal{K}}_{R_1} | m_1, n_1 \kt = \br m_{1}',n_1';0,0|\hat{\mathcal{K}}|m_{1},n_{1};0,0\kt \nonumber \\\
    & = J_{0}(b \, n_1) \;J_{m_1-m_1'}(\alpha_{1}n_{1}') \;\delta_{n_{1}',m_1+n_1}.
\end{align}
\begin{comment}
\begin{figure*}
{\includegraphics[width=75mm, height=50mm]{class_quan_val(N=150,b=0.1).png}}%
{\includegraphics[width=75mm, height=50mm]{class_quan_val(N=60,b=0.1).png}}
\caption{\label{fig:class_quan_val} The eigenvalues of the quantum channel for $N=150$ and $N=60$ with magnitude greater than $0.7$ are plotted on top of the eigenvalues of the classical channel for $\alpha_{1}=2.0$ and $b=0.1$. For the classical channel, $91 \times 91$ modes are used for truncation.}%
\end{figure*}
\end{comment}
The matrix element of the classical channel, $\hat{\mathcal{K}}_{R_1}$ differs from the matrix element of the Koopman associated with a single standard map, $\hat{\mathcal{K}}_1$ in Eq.~(\ref{eq:Koopman1}) just by the term, $J_{0}(b\,n)$. Thus, the nonunitarity of the channel $\hat{\mathcal{K}}_{R_1}$ arises from integrating out the second degree of freedom, as if it were noise.

The quantization of transformation in Eq.~(\ref{eq:standard_map}) gives a unitary, $U = (U_1 \otimes U_2 )U_{b}$. Here, $U_{i}$ has the form as in Eq.~(\ref{eq:quan_stand_map}). In the position basis, the matrix elements are (for example, see 
\cite{Arul_trace,Steven_Deigo2023})
\begin{equation}
\label{eq:quan_unit_pos}
\begin{split}
\bra{n_i'}U_{i}\ket{n_i} = \frac{1}{\sqrt{i N}} \exp\left[ \frac{i \pi }{N}(n_{i}-n_{i}')^2 \right]
\exp\left [\frac{i N\alpha_{i}}{2\pi}\cos\left(2\pi \frac{n_{i}}{N}\right) \right] 
 \end{split}
\end{equation}
\begin{equation}
\label{eq:Ub} 
\begin{split}
\bra{n_{1}'n_{2}'}U_{b} \ket{n_{1}n_{2}}= 
\exp\left [\frac{i N\, b}{2\pi} \cos\left (\frac{2\pi}{N}(n_{1}+n_{2})\right)\right]  \delta_{n_{1},n'_{1}}\delta_{n_{2},n'_{2}}. 
\end{split}
\end{equation}
Unit periodic boundary conditions on $q_i$ and $p_i$ 
has resulted in a finite quantum mechanics of $N$ dimensional Hilbert space where the effective Planck constant is $h=1/N$. Thus the semiclassical limit is $N \rightarrow \infty$. The states have periodic boundary conditions and the value of $N$ is restricted to even integers. The phases are chosen so that  $U$ has both parity and time-reversal symmetry. Such coupled quantum standard maps have been been used to study the effect of chaos on quantum entanglement \cite{Entangling_power_Arul} and the structure of higher-dime.nsional quantum maps \cite{Ritcher_4d}.

The quantum channel in Eq.~(\ref{eq:gen_quan_channel}) for this bipartite unitary is  
\begin{equation}
\label{eq:quan_chan_coup}
\begin{split}
    \mathcal{M}(U)=& \mathcal{M}(U_b) (U_{1}^{\dagger} \otimes U_{1}^{\text{T}}),\;\; \mathcal{M}(U_b)=\frac{1}{N}[U_{b}^{R_{1}\dagger} U_{b}^{R_{1}}]^{R_{2}}\\
    \bra{k'k}\mathcal{M}(U_b)\ket{j'j}&=\frac{1}{N}\sum_{\ell=0}^{N-1} \exp \left\{\frac{i bN}{\pi}\sin[\pi(k-k')/N] \sin[\pi(k+k'-2\ell)/N] \right\} \delta_{kj}\delta_{k'j'}
    \end{split}
\end{equation}
The nonunitary part $\mathcal{M}(U_b)$ is a special case of the Hadamard dephasing channel in Eq.~(\ref{eq:mat_noise_qchannel}) with $\phi_{jk}=(N b/2 \pi) \cos[ 2 \pi (j+k)/N]$. It is diagonal in the position basis used and to a good approximation for large $N$, the sum can be replaced by an integral and we have that 
\beq
\label{eq:MUb}
\bra{k'k}\mathcal{M}(U_b)\ket{j'j}\approx J_0\left[ \frac{b N}{\pi} \sin[\pi (k-k')/N] \right] \delta_{kj}\delta_{k'j'}.
\eeq
Note that while the exact diagonal entries are complex, the approximation is real and valid upto $O(1/N^2)$. The exact value can also be real under a particular boundary condition such as (periodic-antiperiodic) in the two degrees of freedom, but we use the exact complex values in numerical calculations below.
The classical and quantum channels above are independent of the parameter $\alpha_{2}$ of the second degree of freedom. we will refer to $\alpha_1$ as simply $\alpha$. The rest of the paper compares the spectra, both eigenvalues and modes, of the classical channel operator $\hK_{R_1}$  in Eqs.~(\ref{eq:class_chan_uni}) and the quantum channel  $\mathcal{M}(U)$ in (\ref{eq:quan_chan_coup}) for various values of the chaos parameter $\alpha$, interaction strength $b$
and effective (inverse) Planck constant $N$.

\subsection{\label{subsec:Eigval} Eigenvalues: Entangling power and the single ring theorem}
\begin{figure}[h]
\includegraphics[scale=.4]{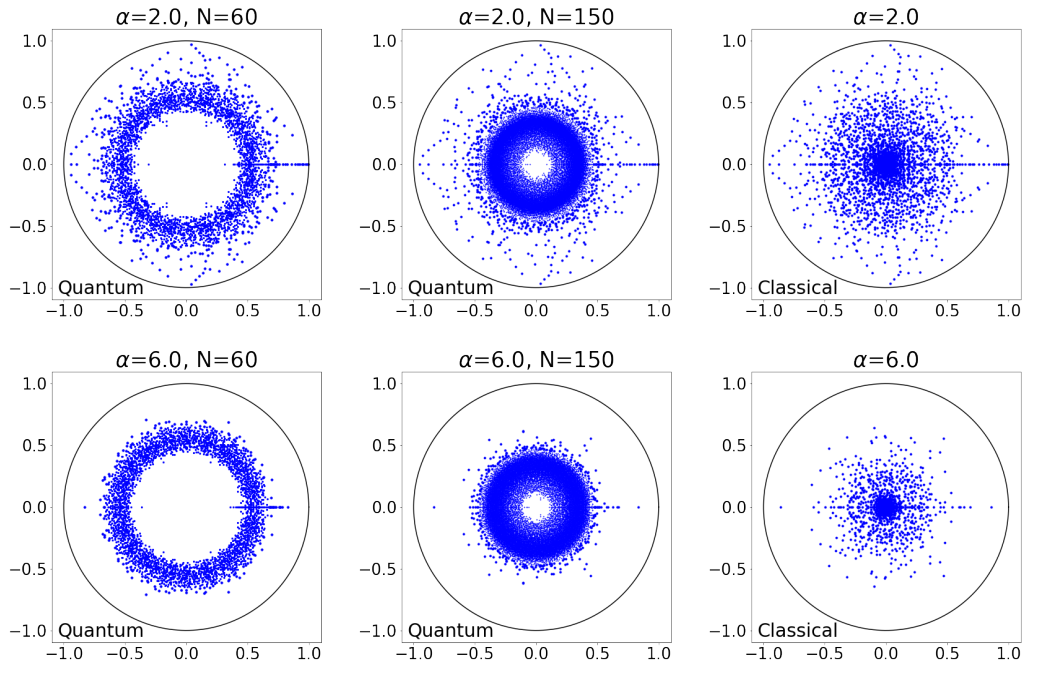}
\caption{\label{fig:eigval} The eigenvalues  $\lambda_i^Q$ from the quantum channel, $\mathcal{M}(U)$ and $\lambda_i^C$ of the classical Koopman channel, $\hat{\mathcal{K}}_{R_1}$, and are shown for two cases of the standard map parameter $\alpha=2.0$ and $6.0$. For the Koopman channel $91 \times 91$ modes are used for the truncation, and for the quantum channel, two different cases of quantization, $N=60$ and $150$ is shown, and the interaction or noise strength is $b=0.1$. The unit circle is show in all the cases as the border.}
\end{figure}

Figure~(\ref{fig:eigval}) shows the eigenvalues of the classical and quantum channels for two different $\alpha$ values, and $N=150 $, and $60$. The interaction or noise strength $b=0.1$, is neither too small nor large in comparison with the $\alpha$ values used. We employ a finite truncation of the classical channel with a total of $91 \times 91$ modes, where the prominent structures have converged. The channels are unital, so there is the trivial eigenvalue $1$. The classical eigenmode in the Fourier basis for the eigenvalue $1$ corresponds to $|0000\kt$, and the quantum eigenmode in the position basis is $|\fullmoon \kt$. In Fig.~(\ref{fig:eigval}), we have plotted all the eigenvalues except the trivial eigenvalue,$1$. Let $|\lambda_{0}|=1\geq |\lambda_{1}| \geq |\lambda_{2}|, .....$ be the eigenvalues of the channels arranged in descending order. From the numerical simulation, it is evident that even for the small values of $b$, $|\lambda_{1}| < 1$. The eigenvalues determine the mixing rates and   $|\lambda_{1}| < 1$ implies that in the thermodynamic limit, all the modes are mixing. 
\begin{table*}
%\hskip-2.5cm
\begin{tabular}{||c | c | c | c | c ||} 
 \hline
$\lambda^{C} $ & $\lambda^{Q} ( N =60) $ & $\lambda^Q ( N = 150 ) $ & $|\lambda^{C}-\lambda^{Q}|$ $( N =60) $ & $|\lambda^{C}-\lambda^{Q}|$($N=150$) \\ [0.5ex] 
 \hline\hline 
0.997283 &	 0.997284 & 0.997282	&	0.000001 &	0.000001 \\ 
 \hline 
 0.988262 & 0.988423 &	0.988282	& 		0.0001600 & 000020  \\
 \hline
0.972954 &	0.973587 & 0.973024 & 0.000634	& 	0.000070	  \\
 \hline
 0.033911-0.965327$i$ &	0.037937-0.969138$i$	& 0.034536-0.965917$i$ & 0.005544	&  0.000860	 \\
 \hline
  0.954322 & 0.954678 &	0.954011 & 0.000357	& 	0.000311	\\ 
 \hline
 -0.952471	& -0.941840	& -0.952129	&	0.010630 &  0.000342  \\
 \hline
0.946673 & 0.944247 &	0.946710 & 0.002425	& 	0.000037	 \\[1ex] 
 \hline
\end{tabular}
\caption{Few large eigenvalues of the classical and quantum channels for $N=150$ and $60$. The difference between the classical and quantum eigenvalues for the two cases ($N=150$ and $60$) are shown in the last two columns, respectively. Here, $\alpha=2.0$, $b=0.1$, and for the classical channel $91 \times 91$ Fourier modes are used for truncation. Six of the 7 modes shown have real eigenvalues.}
\label{table:compare_eval}
\end{table*}

The eigenvalues are either real or come in complex conjugate pairs.  From Fig.~(\ref{fig:eigval}), the outer structure of the eigenvalues of the quantum channel for $N=60 $ and $150$ looks similar to that for a classical channel. 
For a more quantitative comparison,
Table~(\ref{table:compare_eval}) shows some of the large eigenvalues of the classical and quantum channels along with their differences $N=150$ and $N=60$. Preliminary data, not shown, but made plausible by Table~(\ref{table:compare_eval}), indicates that the large eigenvalues such as in the table show the quantum values converge to the classical as a power law $|\lambda_i^Q-\lambda_i^C|\sim N^{-\beta}$, with the power depending on the eigenvalue index as well as other parameters.
As we increase $\alpha$, the dynamics becomes increasingly chaotic,  the eigenvalues move more interior in the unit circle for the same interaction strength, $b$. However the large eigenvalues of the classical channel does seem to converge even in this case as a function of truncation size and the corresponding quantum values converge to the classical from above.

\begin{figure}[h]
\centering
\includegraphics[scale=.5]{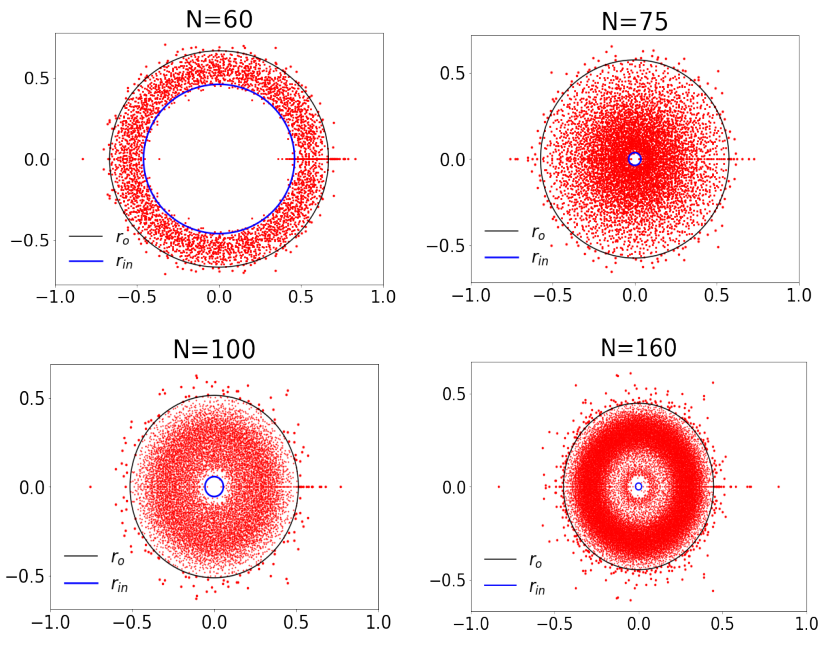}
\caption{\label{fig:annular_radius}The eigenvalues of the quantum channel, $\mathcal{M}(U)$ for $b=0.1$, $\alpha=6.0$, and $N=60,75,100$, and $160$. The black and blue circles are of the outer and inner radius, $r_{o}$ and $r_{in}$ as given by Eq.~(\ref{eq:radius_norm})}
\end{figure}

From Fig.~(\ref{fig:eigval}), we observe that there is a string of significant real eigenvalues for the case of $\alpha=2$ which decreases when the chaos is enhanced at $\alpha=6$. A large number of eigenvalues for the quantum channels though lie in a prominent annular region whose area shrinks with $N$. Quite remarkably this annular region seems to have rotational symmetry and the same size for two quite different values of the parameter $\alpha$, as for $\alpha=2$, there is considerably less chaos than at $\alpha=6$. 
 
An estimate of the inner and outer radius of the annular region for the quantum channels leads us to entangling powers \cite{Ep_Zanardi} and possibly the single-ring theorem \cite{FEINBERG_ZEE,Guionnet2011}. We digress briefly to recall some well-known and not-so-well-known facts relating entanglement and channels.

The entangling power, $e_p(U)$ measures the amount of entanglement generated on an average by an operator, $U$, when acted on an ensemble of pure product states distributed according to Haar 
measure \cite{Ep_Zanardi}. Thus, 
\begin{equation}
    \label{eq:ep_U}
    e_{p}(U) = C_{q} \,\, \overline{ \mathcal{E}(U|\phi_{1}\kt \otimes |\phi_{2}\kt)}^{|\phi_{1}\kt,\phi_{2}\kt}, 
\end{equation}
where,  $\mathcal{E}$ is a measure of entanglement and $C_{q}$ is a scaling factor. Using the scaling as used in \cite{Bernoulli_Suhail}, and the measure of entanglement as linear entropy, we can write the entangling power in terms of linear operator entanglements \cite{Operator_entanglement_Zanardi}, 
\begin{equation}
    \label{eq:ep_opt}
    e_{p}(U) = \frac{1}{E(S)}[E(U)+E(US)-E(S)].
\end{equation}
Here, $S$ is a swap operator and $E(.)$ denotes the operator entanglement, obtained by considering the Schmidt decomposition of the operator in operator Hilbert space. 
\begin{equation}
    \label{eq:opt_ent}
    E(U) = 1 - \frac{1}{N^4} \text{tr}\left[(U^{R_{1}}U^{R_{1}\dagger})^{2}\right], \;\;  E(US) = 1 - \frac{1}{N^4} \text{tr}\left[(U^{T_{2}}U^{T_{2}\dagger})^{2}\right].
\end{equation}
For $S$, the operator entanglement is the maximum possible, $E(S)=1-(1/N^2)$, and $N$ is the dimension of the Hilbert space of states. In fact all dual unitary operators have this maximal value of operator entanglement. The entangling power though is 0 for the swap and can be as large as $1$ for so-called 2-unitary matrices, which exist for all $N>2$ \cite{Goyeneche2015,Suhail_Euler36}.

From the expressions for the channel and the complementary channel in Eq.~(\ref{eq:qchan_act}) and Eq.~(\ref{eq:comp_qchan}), and the operator entanglements, the following connection between the norms of these channels and operator entanglements and entangling power are immediate:
\beq
\begin{split}
&E(U)=1-\frac{1}{N^2} \|\mathcal{M}(U)\|^2, \; 
E(US)=1-\frac{1}{N^2} \|\mathcal{M}_C(U)\|^2,\\ &e_p(U)=\frac{1}{N^2-1}\left(N^2+1-\|\mathcal{M}(U)\|^2-\|\mathcal{M}_C(U)\|^2 \right).
\end{split}
\eeq
Here $\|A\|=\sqrt{\tr(A A^{\dagger})}$ is the Hilbert-Schmidt or Frobenius operator norm.
The relation with entangling power can be written more compactly as 
\beq
(N^2-1)(1-e_p(U))=
\|\Tilde{\mathcal{M}}(U)\|^2+\|\Tilde{\mathcal{M}}_C(U)\|^2 
\eeq
where $\|\Tilde{\mathcal{M}}(U)\|^2 =\|\mathcal{M}(U)\|^2-1$, and similarly for the complementary channel, the tilde channels result from projecting out the trivial mode with eigenvalue 1. If $U$ is dual-unitary, $\|\Tilde{\mathcal{M}}(U)\|^2 =0$, the complementary channel carries the weight of the entangling power, and if $U$ is T-dual, $\|\Tilde{\mathcal{M}}_C(U)\|^2=0$, and the channel alone is active.

Returning to the coupled standard map under consideration $U = (U_{1} \otimes U_{2})\,U_{b} $, and $e_p(U)=e_p(U_b)$, $E(U)=E(U_{b})$ and $E(US)=E(U_bS)$, as they are all local unitary invariant. As $U_{b}$ is diagonal, its partial transpose is itself, it is T-dual, and we have $E(US)= E(S)$. Therefore, the entangling power is $e_p(U)=E(U)/E(S)$, the operator entanglement fraction in relation to the swap and $(N^2-1)(1-e_p(U))=
\|\Tilde{\mathcal{M}}(U)\|^2$, as perhaps first observed in \cite{Bernoulli_Suhail}.
Using Eq.~(\ref{eq:MUb}), and the connections noted above, the entangling power $e_p(U)=e_p(U_b)$ can be expressed in terms of Bessel's functions for large $N$ as
\begin{equation}
    \label{eq:std_ep}
    e_p(U_b) = 1-2 \int_0^1 x J_0^2\left(\frac{bN}{\pi} \sin \pi x \right) \, dx + O(1/N^2)=1-\frac{C}{bN}+O(1/N^2),
\end{equation}
where $C$, in addition to constants,  contains oscillatory and logarithmic contributions (in $bN$) originating from the
asymptotics of the Bessel function.

Note that Eq.~(\ref{eq:MUb}) implies that the singular values of the channel $\mathcal{M}(U)$ are approximately,
\beq
\sigma_{kk'}=\left|J_0\left[\frac{bN}{\pi}\sin \frac{\pi (k-k')}{N} \right]\right|,
\eeq
and are (unrelated to the large $N$ approximation) not dependent on the standard map parameter $\alpha$. Whereas many of the large magnitude eigenvalues of the channels as shown in Fig.~(\ref{fig:eigval}) do depend on $\alpha$. As $\mathcal{M}(U)=\mathcal{M}(U_b)(U_1^{\dagger} \otimes U^*_1)$, considering $U_1$ as being sampled from the random Haar measure (Circular Unitary Ensemble, CUE), it was shown analytically in \cite{Bernoulli_Suhail} that approximately the expected (absolute) value of the largest nontrivial eigenvalues of $\mathcal{M}(U)$ will be $\mathbbm{E}( |\lambda_1|) = \sqrt{1-e_p(U)}=\sqrt{1-e_p(U_b)}$. Averaging over the local unitaries has washed away any outliers and the principal eigenvalue depends only on the interaction. Indeed averaging over local unitaries produces a local unitary invariant. We notice that this is too extreme and throws out the dynamical baby with the bathwater of randomness. 

However, this estimate does give us the annular region's outer radius as
\begin{equation}
    r_{o} = \sqrt{1-e_{p}(U_b)} \approx \sqrt{C/bN},
    \label{eq:ro}
\end{equation}
 dependent only on the interaction between the system and the ``environment" and oblivious to the local dynamics of the system. Figure.~(\ref{fig:annular_radius}) 
 shows a circle of radius $r_o$, determined from the entangling power, and as we change $N$ we see how the annular region shrinks. Fig.~(\ref{fig:eigval}) also shows the independence of the outer annular radius with change in the local parameter $\alpha$.

The annular regions noted above are reminiscent of the ``single-ring theorem" \cite{FEINBERG_ZEE, Guionnet2011}.
This deals with the general question of the distribution of eigenvalues of a non-Hermitian matrix ensemble with a deterministic or known singular value distribution. It was found, rather surprisingly that the eigenvalues populate only a single rotationally invariant annular region, the single ring. In \cite{Guionnet2011}, the authors have proved the single-ring theorem for the  matrix ensemble $M_n = U_n T_n V_n$, where $U_n$ and $V_n$ are independent Haar random unitaries from the group $\mathcal{U}(n)$, and $T_n$ is a diagonal matrix with real entries. The quantum channel $\mathcal{M}(U)$, say from Eq.~(\ref{eq:quan_chan_coup})
is not exactly of this form, but is also not without similarities. As $M_n$ and $T_n V_n U_n^{\dagger}$ have the same eigenvalues, we see that the analogy is that $T_n$ is the (to a good approximation) real and diagonal $\mathcal{M}(U_b)$, which determines the singular values of the channel. While the single ring theorem assumes strong statistical properties on the unitaries $U_n$ and $V_n$, in this case they are determinsistic, being quantum standard maps and more importantly having a particular tensor product structure. The quantum standard map for large $\alpha$ can be a proxy for the CUE (actually in this case the orthogonal ensemble as it respects time-reversal), however there is also a particular product structure. 

Nevertheless, the presence of a prominent annular region that is essentially interaction dependent, motivates us to compare with the known results from the single ring theorem, such as the inner and outer radius for the annular region \cite{Feinberd_Zee_Scalettar_radius}. These are expressed in terms of the singular values of $T_n$ or the eigenvalue distribution of $T_n^2$. After some translation to the notation in this paper, their result is

    \begin{equation}
    \label{eq:radius_norm}
        r_{o}^2 =\frac{\overline{\|\Tilde{\mathcal{M}}(U)\|^2}}{N^2-1} = {1-e_{p}(U)}, \;\;
       \frac{1}{r_{in}^2} =\frac{\overline{\|\Tilde{\mathcal{M}}^{-1}(U)\|^2}}{N^2-1}.
    \end{equation}
    The overbar accommodates a nondeterministic diagonal, but is irrelevant to considerations here and may be ignored. Quite remarkably (and mildly amusingly) this estimate of the outer radius matches the result for channels in \cite{Bernoulli_Suhail}, and recounted in Eq.~(\ref{eq:ro}). The bonus is the inner radius which is determined from the inverse of the channel. Note that if there are any vanishing singular values, indeed $r_{in}=0$, as would be the case for dual unitary channels. When compared with the quantum channel, see Fig.~(\ref{fig:annular_radius}), we find good agreement with these estimates but also  deviations for the inner radius for large $N$. Numerical results then indicate a broader setting for the single-ring theorem and it's relevance to dephasing quantum channels. We note that in a random matrix model of channels the appearance of a single ring has already appeared \cite{Lucas_Prosen_singlering}.

\subsection{\label{sec:level6} Eigenvalues:  the second largest eigenvalue}

In determining the degree of mixing or the mixing rates of the channels the magnitude of the largest non-trivial eigenvalue $|\lambda_{1}|$ plays an important role.
%\cite{Bernoulli_Suhail},\cite{Ergodic_channels_Aravinda}. 
In the setting of many-body systems, it has been shown \cite{Dual_Prosen} that the two-point correlation along the light cone for dual unitary circuits can be derived from ${\lambda_{i}}$, and if $|\lambda_{1}| < 1$ the correlation function decays. The decay rate of the two-point correlation in the thermodynamic limit is determined from the magnitude of the largest non-trivial eigenvalue, $\lambda_{1}$. In this part of the paper we drop the superscript to indicate quantum or classical on the eigenvalues as they share similar features. The context will make it clear if we are referring to any particular one alone.

\begin{figure}[h]
\includegraphics[scale=.5]{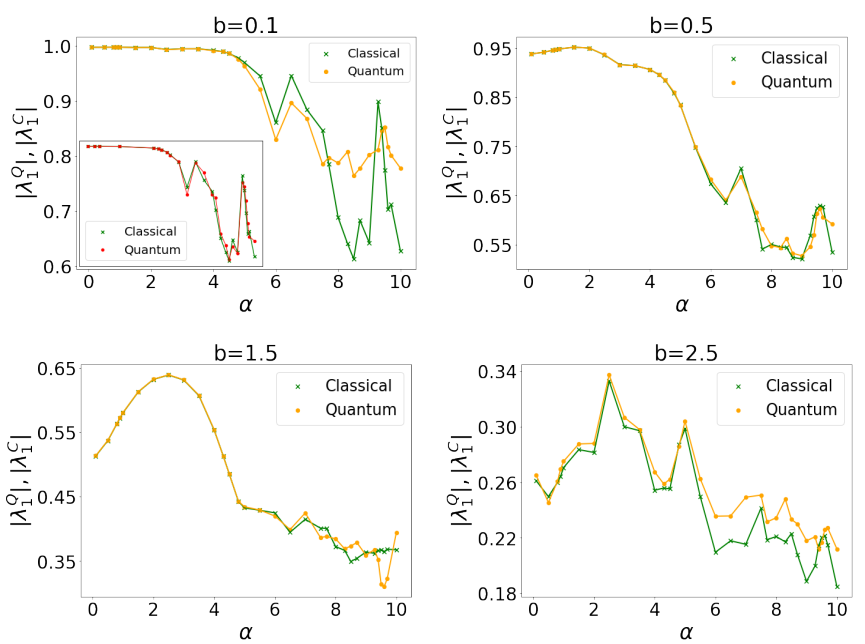}
\caption{\label{fig:large_K} Variation of largest non-trivial eigenvalue with $\alpha$ for $b=0.1 , 0.5 , 1.5, $ and $2.5$ for the classical channel ($91 \times 91 $ modes) and quantum channel ($N=60$). The inset(in the upper left panel) illustrates the results for both channels but with the quantum case $N = 150$.}%
\end{figure}

For the standard map channels under consideration in this paper, we observe that as $\alpha$ increases and the dynamics transitions from a near-integrable to a chaotic regime, $\lambda_1$ is pushed more to the interior of the unit circle but, interestingly, not monotonically. The variation of $|\lambda_{1}|$ with $\alpha$ for different interaction strengths is shown in Fig.~(\ref{fig:large_K}). 
 
In a work by one of us \cite{Dual_Koopman_Arul}, it has been stated that for the integrable case when $\alpha=0$, some of the eigenvalues of the classical channel $\hat{\mathcal{K}}_{R_1}$ are $J_{0}(b\,k)$ with the corresponding eigenmode  $\br m,n|\phi_{k}\kt = \delta_{m,0}\delta_{n,k}$ for $k=0,1,\cdots$. Indeed, the numerical value of $\lambda_{1}$ matches well with $J_{0}(b)$ for the classical channel, and even for the quantum, see Table~(\ref{table:large_eval_jac}). 

\begin{table}[h]
\centering
\begin{tabular}{||c | c | c | c ||} 
\hline
$b$ & $\lambda_{1}^{C} $ &  $\lambda_{1}^{Q} $ & $J_{0}(b)$ \\ [0.5ex] 
 \hline\hline 
0.1 & 0.997501562 &	0.9975038	& 0.99750156  \\ 
 \hline 
0.5 & 0.93846981  &	0.9385251	&  0.93846981  \\
 \hline
1.0 & 0.76519769 &	0.7653987	&  0.76519769  \\
 \hline
1.5& 0.51182767 &	0.51220999	&  0.51182767 \\[1ex] 
 \hline
\end{tabular}
%\end{table}
%\begin{table}
\begin{tabular}{||c | c | c | c ||} 
 \hline
$b$ & $\lambda_{1}^{C} $ &  $\lambda_{1}^{Q} $ & $J_{0}(b)$ \\ [0.5ex] 
 \hline\hline 
0.1 & 0.99750784 &	0.99751011	& 0.99750156  \\ 
 \hline 
0.5 & 0.93862337 &	0.93867842	&  0.93846981  \\
 \hline
1.0 & 0.76578154 &	0.76598152	&  0.76519769  \\
 \hline
1.5& 0.5130379 &	0.51341805	&  0.511827672 \\ [1ex] 
 \hline
\end{tabular}
%\captionsetup{position=bottom}
\caption{Largest eigenvalue of the quantum and classical channel for different interaction strength, $b$ and comparing it with zeroth order Bessels's function, $J_{0}(b)$ for $\alpha =0$ (left), and $\alpha=0.1$ (right). The eigenvalues of the quantum channel are for $N=60$, and for the classical channel, $91 \times 91 $ modes are used for truncation.}
\label{table:large_eval_jac}
\end{table}

 There is a slight decrease from $J_0(b)$ in the value of $|\lambda_1|$ as $\alpha>0$, and is in fact real. The significant drop in $|\lambda_{1}|$ around $\alpha_{1} \approx 4.0 $ is correlated to the fixed point at  $(q,p)=(0,0)$ losing stability for the ``clean" standard map at $\alpha=4$. Thus, the change of stability of the periodic orbit at $(q,p)=(0,0)$ is also being captured by the $|\lambda_{1}|$. This is further validated ahead when we examine the eigenmodes. Increasing $\alpha$ further reveals features that are present both in the classical and quantum channels. The value of $|\lambda_1|$ increases at two distinct points where $\alpha\approx 6.5$, and $\alpha \approx 9.3$. These can be correlated to the appearance of further stable fixed points due to bifurcations. We have already mentioned the accelerator modes that appear in an interval after $\alpha=2 \pi$, and the local maximum around $\alpha=6.5$ in Fig.~(\ref{fig:large_K}) can be correlated to this. The second very prominent and sharp peak around $\alpha=9.3$ is a point where small regular islands are forming and have been detected using finite time Lyapunov or stability/Lyapunov exponents (FTSE/ FTLE) \cite{FTSE_Arul}. The complex nature of the standard map with a growing average Lyapunov exponent, but with the likelihood of small islands also being present is reflected in this plot. Thus the variation of $|\lambda_{1}|$ can help us detect both small and large islands in the classical phase space. The inset in the panel corresponding to $b=0.1$, shows the results when the classical limit is approached further, with $N=150$.
Notice the extremely good classical-quantum correspondence in this, as both follow even small variations faithfully,
and the differences that were seen for $N=60$ are significantly diminshed.
classical

When the coupling parameter, or noise strength is increased to $b=0.5$,
we can still notice these features in a rounded and less contrasting manner in Fig.~(\ref{fig:large_K}). The loss of stability of the fixed points, the accelerator modes and the bifurcation around 9.3 are still visible. When $b$ is increased further to $b=1.5$, the noise has now significantly pulled down the value of $|\lambda_1|$, but interestingly it shows a prominent nonmonotonic behaviour even in the interval $[0,4]$. It is possible that the growing nonlinear resonance region of the fixed point at the origin provides room for the noise to be less destructive at around $\alpha=2$. Further, for large values such as $b= 2.5$, the $\lambda_{1}$ seems to fluctuate somewhat randomly as we vary $\alpha$ and the correspondence with the clean standard map is more or less lost.

\section{\label{sec:Modes}Eigenmodes: Classical-Quantum correspondence and scarring}

Eigenmodes are naturally both interesting and important. In the unitary closed case, the study of eigenfunctions has revealed a somewhat bewildering variety of patterns. For a fully chaotic system, it was expected that the eigenstates be similar to interference patterns of  random waves \cite{Berry_1977_reg}. In particular the Wigner function of eigenstates was expected, under local averaging, to result in a microcanonical distribution on an energy shell. However, many of them were found to be strongly imprinted by patterns that could be traced back to classical isolated and unstable periodic orbits \cite{Heller} and also their associated homoclinic and heteroclinic excursions \cite{Baker_Saraceno,KAPLAN_Heller1998}. Semiclassical and Gaussian wavepacket based theories exist for such scarring, at least in a spectral averaged sense \cite{Berry1989_scar,BOGOMOLNY1988169,KAPLAN_Heller1998}. 

As the largest non-trivial eigenvalue of the channel, $\lambda_{1}$, seems to capture information about classical phase space structures, it is natural to first examine the corresponding eigenmodes of the channel. 
%Let $|\mathcal{X}_{c}^{R}\kt$ be a eigenmode associated with a complex eigenvalue, $\lambda_{c}$ then the eigenmode associated with $\lambda_{c}^{*}$ is $|\mathcal{X}_{c}^{R\dagger}\kt$.
For an operator $A$, the Husimi representation, which is a Gaussian smoothed Wigner function is given by 
    $h_{A}(q,p)= |\br pq |A| pq \kt |^{2}$.
    Here the state $\ket{pq}$ is a coherent state centered at momentum $p$ and position $q$. As we have a toral topology we will use the coherent states on the torus as described by Saraceno in \cite{Baker_Saraceno}. 
We  can define the Husimi representation of the right eigenmode of the quantum channel, $\mathcal{M}(U)$ as 
\begin{equation}
\label{eq:husimi_mode}
    h_{\mathcal{X}_{R}}(q,p)= |\br pq| \otimes \br pq^* | \mathcal{X}^{R}\kt |^{2}.
\end{equation}
 As the classical channel was represented in Fourier basis, in this case we simply make of change of basis from $(m,n)$ to $(q,p)$. 

\begin{figure}
\centering
\includegraphics[scale=.7]{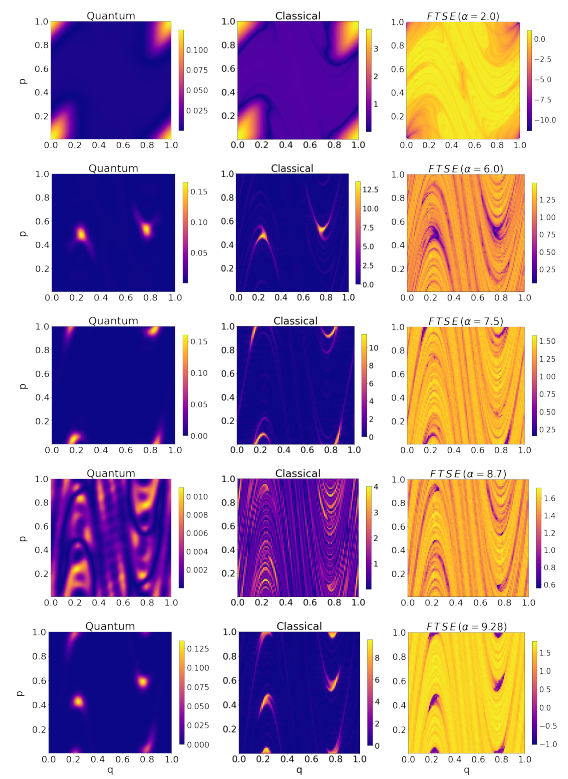}
\caption{\label{fig:largest_mode}Eigenmode of the quantum and classical channel associated with the largest non-trivial eigenvalue for $b=0.1$ is plotted alongside the FTSE for a single standard map at $\alpha=2.0, 6.0, 7.5, 8.7 $, and $9.28$ Here, $N=100$, and for the classical channel $91 \times 91$ modes are used for truncation.}
\end{figure}

Fig.~(\ref{fig:largest_mode}) shows the eigenmodes of the quantum and classical channel plotted alongside the average finite time stability exponent(FTSE). The finite time stability exponent measures the local stability around a phase space point and can be approximated as \cite{FTSE_Arul}
\begin{equation}
\label{eq:ftse}
    \lambda_{t}(q_0,p_0) \approx \frac{1}{t} \text{ln}|\text{Tr}[M_{t}(q_0,p_0]|
\end{equation}
where $M_t$ is the time-$t$ Jacobian or stability matrix, which is a product of the stability matrices along an orbit that originates at $(q_0,p_0)$: 
\begin{equation}
    \label{eq:stability_mat}
    M_{t}(q_{0},p_0) = \prod_{i=0}^{t-1}  M(q_{i}) =  \prod_{i=0}^{t-1}\begin{pmatrix}
1 -V''(q_{i}) &  1 \\
-V''(q_i) & 1 
\end{pmatrix}, 
\end{equation}
where $V(q) = -(\alpha/4\pi^{2}) \cos(2\pi q)$. The phase space is coarse grained into a fine grid and the average of $\lambda_t$ is taken as the value at the center of a cell.

Figure~(\ref{fig:largest_mode}) shows the both the classical and quantum channel eigenmodes corresponding to the largest nontrivial eigenvalue for 5 different values of the standard map parameter $\alpha$, while the interaction/noise parameter $b=0.1$ is quite small. Shown in the third column is the classical map's FTSE, rather than just the orbits. Regions that are darker correspond to more stable regions. Firstly, the classical-quantum correspondence is remarkable; the classical Koopman channel modes have sharper features and the unstable manifolds of nearby short unstable periodic points are visible. 
In more detail, the case of $\alpha=2.0$
is dominated by the stable fixed point region around the origin. It is quite surprising that even small (in comparison with $h=1/N$), but stable, islands at $\alpha=6.0$, and $7.5$ are being captured by the modes. At $\alpha=8.7$ there does not appear to be any significant regular region, and the most significant eigenmode is seen to be delocalized. The similarity between the classical Koopman channel mode and the FTSE which shows the foliations of the unstable manifolds is striking. The quantum mode is also delocalized and is a grainier version, taking into account the finite $\hbar$. 

We note that when $b=0$, and the channel is clean and unitary, there is no ordering of the eigenstates on the unit circle according to the localization of the states. However, introducing a small noise is sufficient to distinguish the localized states and make them the most significant. It is also plausible that the prominent features present in the modes are really the unstable manifolds that are highly dense around stable regions; this will become more plausible when we increase the noise strength $b$ further below.

\begin{figure}[h]
    \centering
    \includegraphics[width=135mm,height=106mm]{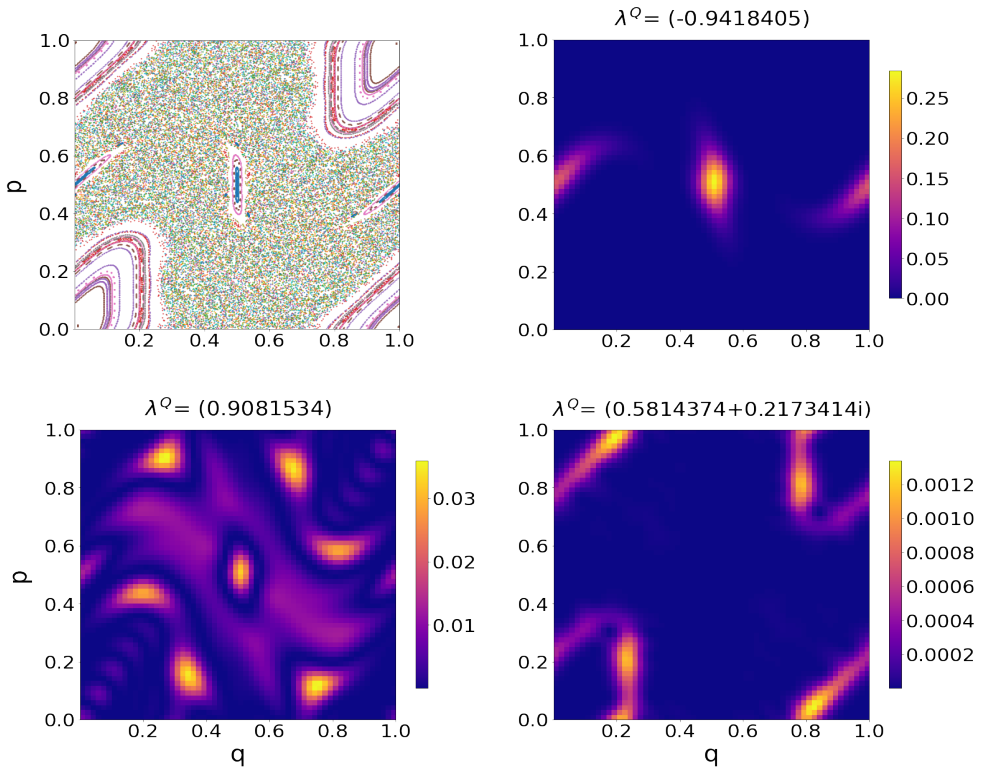}
    \caption{The Husimi function of 3 eigenmodes of the quantum channel for $N=60$, $\alpha=2.0$, and $b=0.1$ showing the imprints of stable period-2 orbit, small resonance chain of length 6 and a tori, all visible in the classical phase space (upper left).}
    \label{fig:quan_mode}
\end{figure}

 Significant modes, other than those corresponding to the largest nontrivial eigenvalue, are also found to have various features of the phase space. For instance, when $\alpha=2.0$ the classical phase space  is that of a typical mixed phase space.
Husimi of a few modes of the quantum channel is plotted, along with the classical phase space, in Fig.~(\ref{fig:quan_mode}) for this case with $b=0.1$.  The stable period two orbit,$(0,0.5), (0.5,0.5)$, and the island around $(q,p)=(0,0)$ seem to imprint many of the eigenmodes. We also see a  length 6 chain of small resonance islands along with the period two appearing in the lower left figure,
while the last is influenced by a quasiperiodic tori.

\begin{figure}[h]
    \centering
    \hspace*{-1.4cm}   \includegraphics[width=195mm,height=112mm]{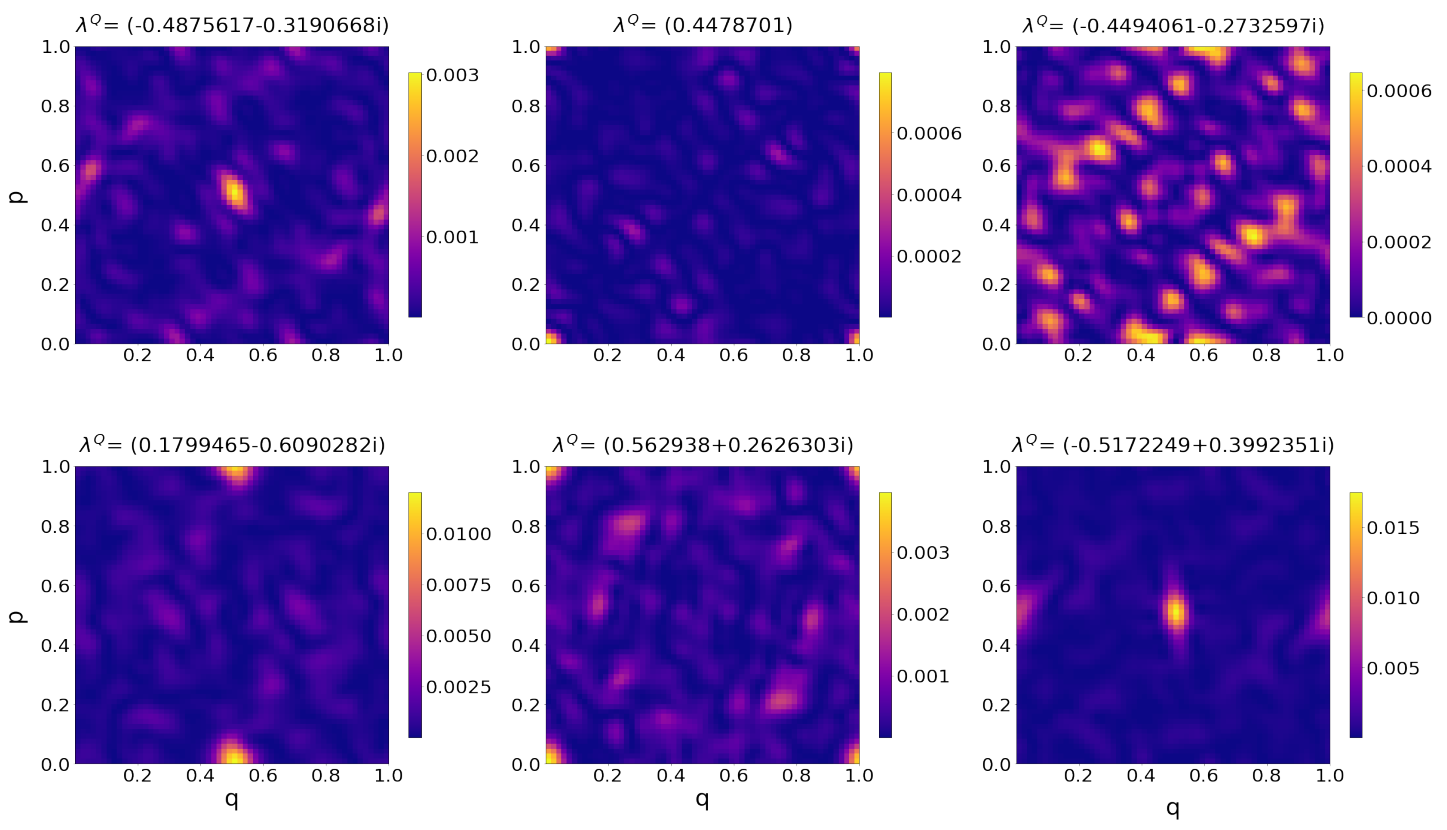}
    \caption{The Husimi function of six eigenmodes of the quantum channel for $N=60$, $b=0.1$, $\alpha=6.0$ (first row) and $\alpha=7.5$ (second row). The associated classical phase space is shown in Fig.~(\ref{fig:hus_unit_state}). 
Unstable fixed points and period-2 orbits are seen to scar the eigenmodes, except for the upper-rightmost case that seems dominated by the chaotic sea.}
    \label{fig:quan1_mode6.0}
\end{figure}

\begin{figure}[h]
\centering
\includegraphics[scale=.4]{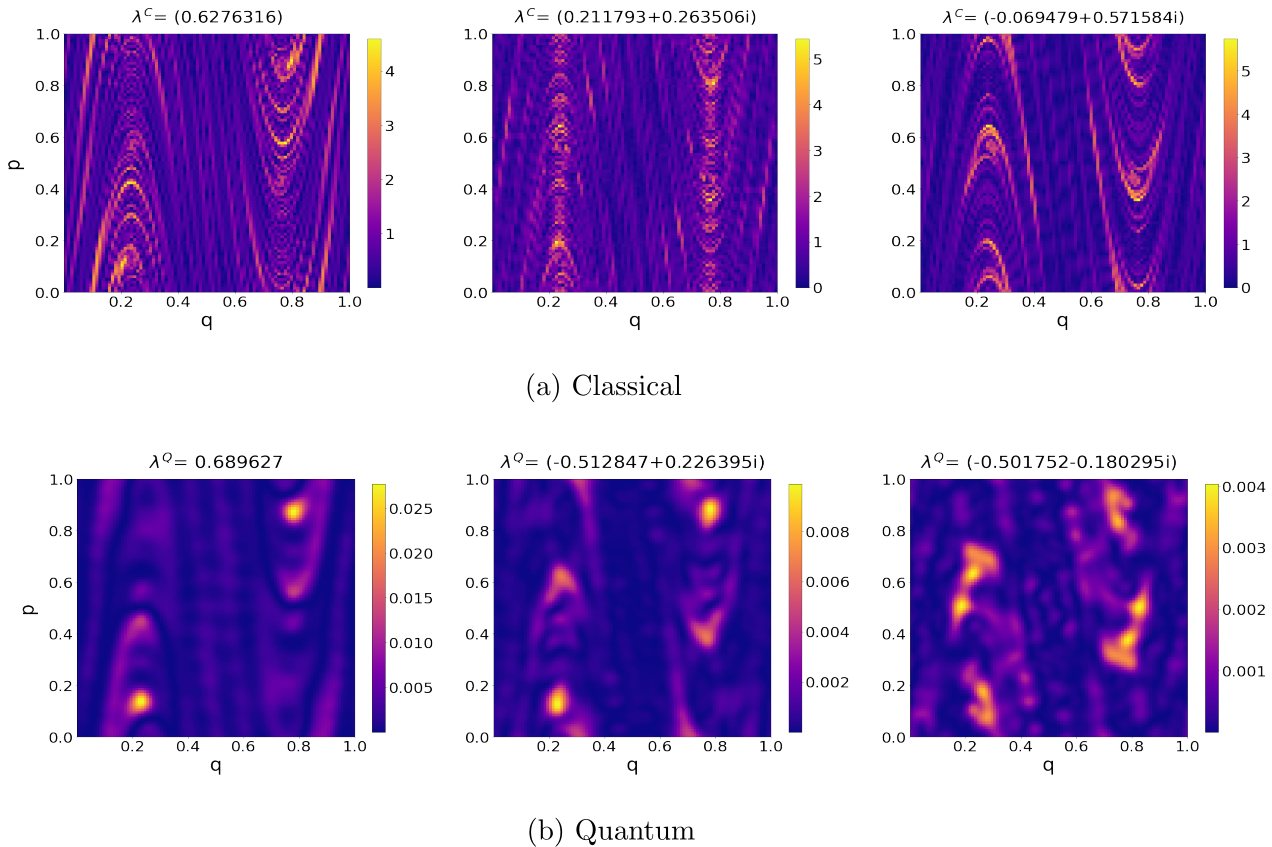}
\caption{\label{fig:mode_10.0} A few eigenmodes of (a) Classical and (b) Quantum channel for $\alpha = 10.0$. The noise strength, $b=0.1$, $N=100$ for the quantum channel, and $91 \times 91$ Fourier modes are used for truncation of the classical channel.}
\end{figure}

\begin{figure}
\centering
\includegraphics[scale=.3]{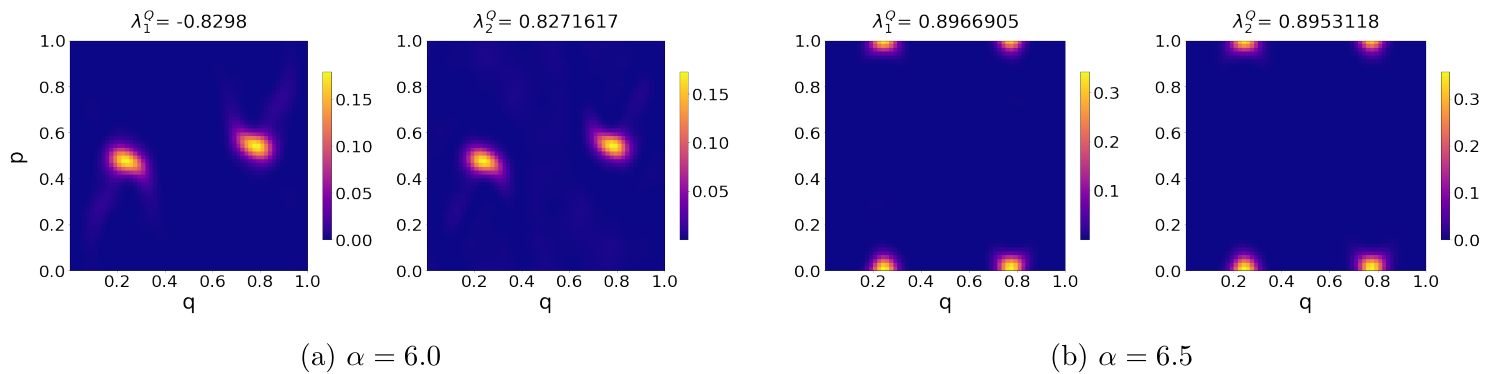}
\caption{\label{fig:mode_tunnel}The Husimi function of two eigenmodes of the quantum channel corresponding to the largest and second-largest non-trivial eigenvalue, $\lambda_{1}^{Q}$ and $\lambda_{2}^{Q}$. Here $N=100$, $b=0.1$, and (a) $\alpha = 6.0$, 
(b) $\alpha=6.5$. }
\end{figure}

\begin{figure}
\centering
\includegraphics[scale=.35]{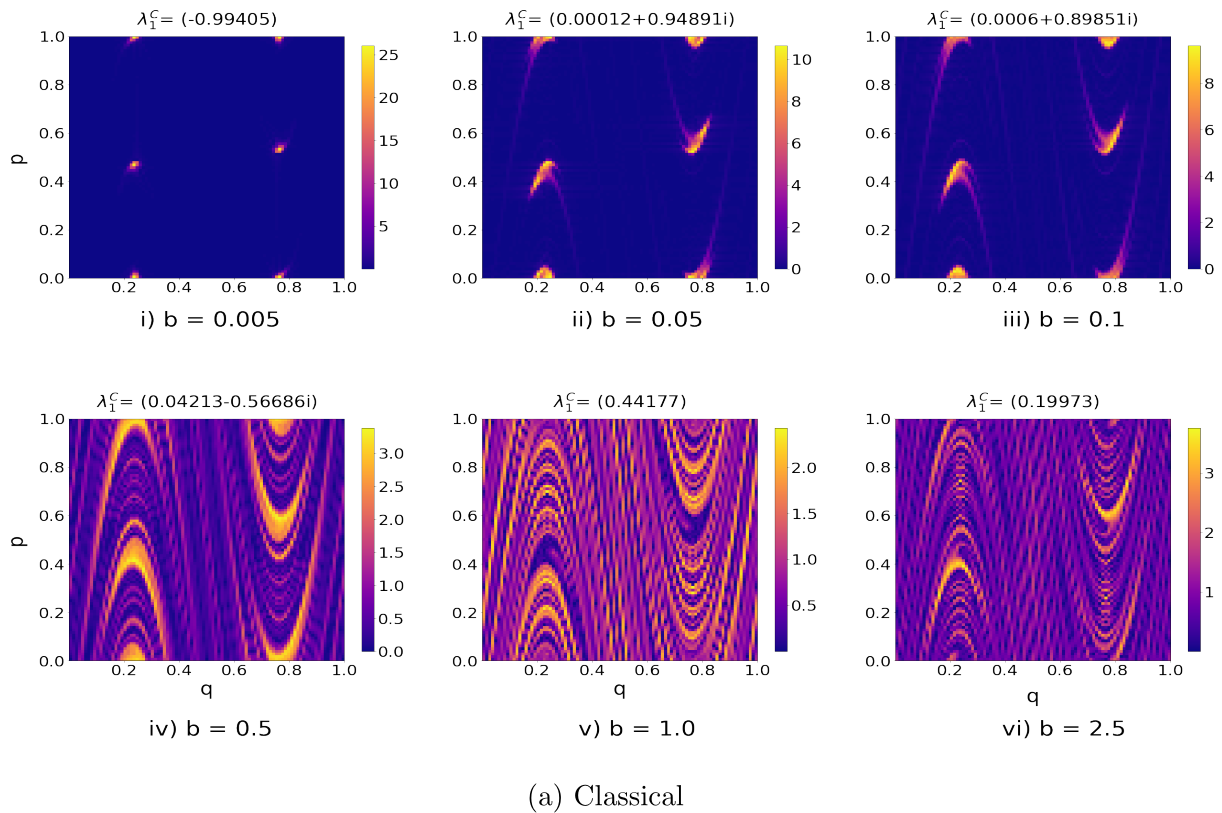}
\includegraphics[scale=.35]{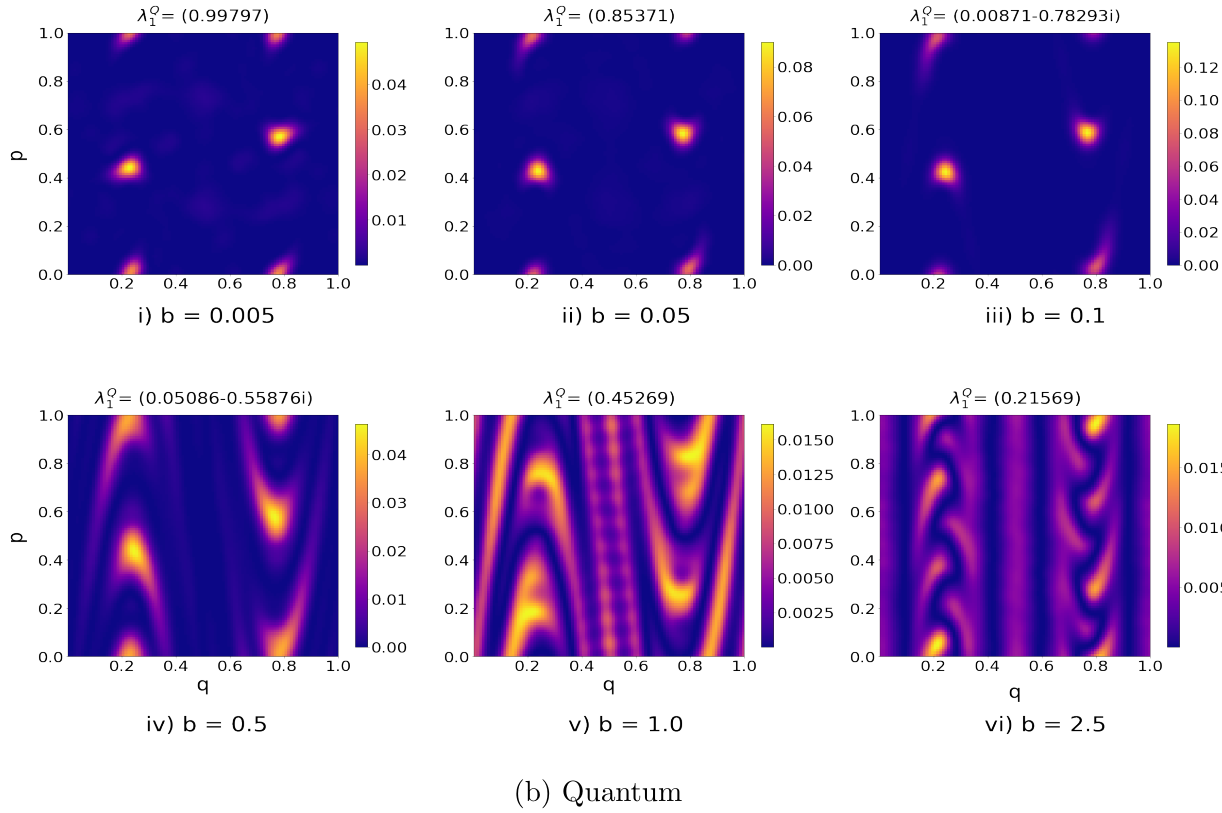}
\caption{\label{fig:noise_mode}The largest (in absolute magnitude) non-trivial eigenmode of the a) Classical channel, and b) Quantum channel for different noise strengths. Here, $\alpha =9.28$, $N=100$, and $91 \times 91$ modes are used for truncation of the classical channel. }
\end{figure}

A quantum scar or ``scarring" is usually referred to as the enhanced probability of a quantum state around an unstable periodic orbit \cite{Heller,KAPLAN_Heller1998}.  The channel eigenmodes have enhanced probability both around stable and unstable orbits. Figure~(\ref{fig:quan1_mode6.0}) shows scarring at $\alpha=6.0$ and $7.5$ by what are now unstable fixed points at $(q,p)=(0,0)$,$(q,p)=(0.5,0)$,and by period two orbit at $(q,p)=(0,0.5),\, (q,p)=(0.5,0.5)$. The associated classical phase space is shown in Fig.~(\ref{fig:hus_unit_state}). Classical channels also show such scarring features. At $\alpha=10.0$, the single standard map does not seem to have any stable islands (although this is not rigorously proven). For this case, Fig.~(\ref{fig:mode_10.0}) displays a selection of 3 classical and 3 quantum modes that have large eigenvalues.
Without the crutches of stable islands the modes spread out along the unstable manifold and selectively highlight parts of it, such as the FTSE does. Note that we have verified that the largest nontrivial eigenvalue in this case $\approx 0.627$ seems quite converged and does not change significantly with truncation size. The quantum modes seem blurred versions of the classical modes, again three significant ones have been displayed.
We note that recently  \cite{Scarring_noise_Yoshida}, scarring in the eigenmodes of a noisy classical Perron-Frobenius operator has been reported. 

Unstable fixed points and period-2 orbits are seen to scar the eigenmodes, except for the upper-rightmost case that seems dominated by the chaotic sea.
The phenomenon of chaos assisted tunneling has been studied in low-dimensional systems \cite{Steven_tunneling,Keshavamurthy2011}, and the influence of noise and environment must be of interest. In this context,  Fig.~(\ref{fig:mode_tunnel}) shows the largest and the second largest non-trivial eigenvalue of the quantum channel for $\alpha=6.0$ and $\alpha=6.5$. The difference between the two modes is that at $\alpha=6$, the stable but small period two orbit shows up, and the other is scarred by a fixed point. For the fixed point, the eigenvalues are nearly degenerate while, for the period two orbit they are nearly $\pi$ apart, and hence the eigenvalues of the square of the channel is nearly degenerate in this case.  The figure is for $b=0.1$. We do have twin partners for other $b$ values but the eigenvalues are not near degenerate or $\pi$ apart, and this level of degeneracy is attained by tuning $b$. Thus we present this as a very preliminary motivation for the study of chaos assisted tunneling in general channels.

For a given $\alpha\, (=9.28)$, where apart from small periodic chains, the dynamics is completely chaotic, the phase space representation of the largest mode of both the classical and quantum channels, are shown in Fig.~(\ref{fig:noise_mode}), for various values of the noise parameter $b$. It is seen in the classical case, that the unstable manifold lights up increasingly more as the noise or interaction increases. The appearance is similar to  that of the classical FTSE, lighting up the regions of relatively high stability. The structural integrity of this manifold seems quite robust, in the sense that even with as large as value for $b$ as $2.5$, the manifolds ``appear" similar to that of the single standard map. The quantum case however shows very different interference patters for large $b$ where braid like structures appear. It is clear that quantum-classical correspondence breaks down at this value of $N$ and large values of $b$.

\subsection{A Hermitian submatrix}

Many of the large eigenvalues of the channel $\mathcal{M}(U)$ are real, especially for smaller values of the parameter $b$. The number of real eigenvalues and their evolution with the parameters is of interest.  When $b=0$ the quantum channel, $\mathcal{M}(U)=U_{1}^{\dagger} \otimes U_{1}^{\text{T}}$,
and if $ U_{1}|\phi_{j}\kt = \exp(i\phi_{j})|\phi_{j}\kt $ are the eigenvalue equations, 
the eigenmodes of $\mathcal{M}(U)$ are $|\phi_{j} \kt | \phi_{k}^* \kt $, with eigenvalues $\exp(-i (\phi_j-\phi_k))$. The $N$ dimensional degenerate subspace with $j=k$,
where the eigenvalues are all $1$ is of interest. As the noise is switched on, only one of the eigenvalues stays $1.0$, while the rest move inside depending on $b$ and $\alpha$. Of special interest to us is the largest non-trivial eigenvalue and the eigenmode associated with it in terms of the eigenmodes of the unitary operator. 

For $b\neq 0$, the channel $M(U)=\mathcal{M}_{b}(U_b)(U_{1}^{\dagger} \otimes U_{1}^{\text{T}})$
where
\begin{equation}
    \label{eq:mat_db}
    \br i j |\mathcal{M}_{b}(U_b)|k l \kt = d_{i j}(b) \delta_{ik} \delta_{jl},\;\;d_{ij}(b) = \frac{1}{N} \sum_{k=1}^{N} \exp[i(\theta_{ik}-\theta_{jk})],
\end{equation}
is a diagonal matrix, that is the nonunitary part. It has the symmetry that $d_{ij}=d_{ji}^*$.

The matrix element of the channel in the free  basis, $\{|\phi_{i}\kt |\phi_{j}^*\kt\}$ is 
\begin{equation}
    \label{eq:qchann_ubasis}
     \br \phi_{k'}\phi_{l'}^*|\mathcal{M}(U)|\phi_{k}\phi_{l}^*\kt = e^{-i(\phi_{k}-\phi_{l})}
    \sum_{ij} d_{ij}(b) \br \phi_{k'}|i\kt \br j|\phi_{l'} \kt \br i|\phi_k\kt \br \phi_l|j \kt.
\end{equation}

\begin{figure}
\centering
\includegraphics[scale=.4]{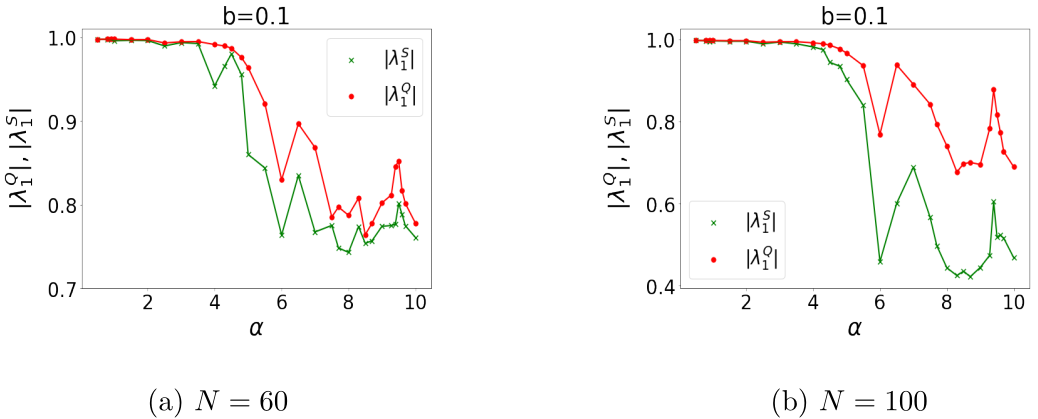}
\caption{The variation of the largest eigenvalues of the quantum channel, $\mathcal{M}(U)$ and the submatrix, $\mathcal{M}_{S}$ with $\alpha$ for $b=0.1$, and a) $N=60$, b)$N=100$. }
\label{fig:eigval_submat_chan}
\end{figure}
Considering only the subspace for which $k'=l'$ and $k=l$, which in the $b=0$ limit is the degenerate space with eigenvalue $1$, we get a submatrix, $\mathcal{M}_{S}$ of order $N$ with elements
\begin{equation}
    \label{eq:submat}
    \br \phi_{k'}\phi_{k'}^* | \mathcal{M}_{S} |\phi_{k}\phi_{k}^* \kt = 
     \sum_{i j} d_{ij}(b)  \br \phi_{k'}|i\kt \br j|\phi_{k'} \kt \br i|\phi_k\kt \br \phi_k|j\kt.
\end{equation}

\begin{figure}
\centering
\includegraphics[scale=.3]{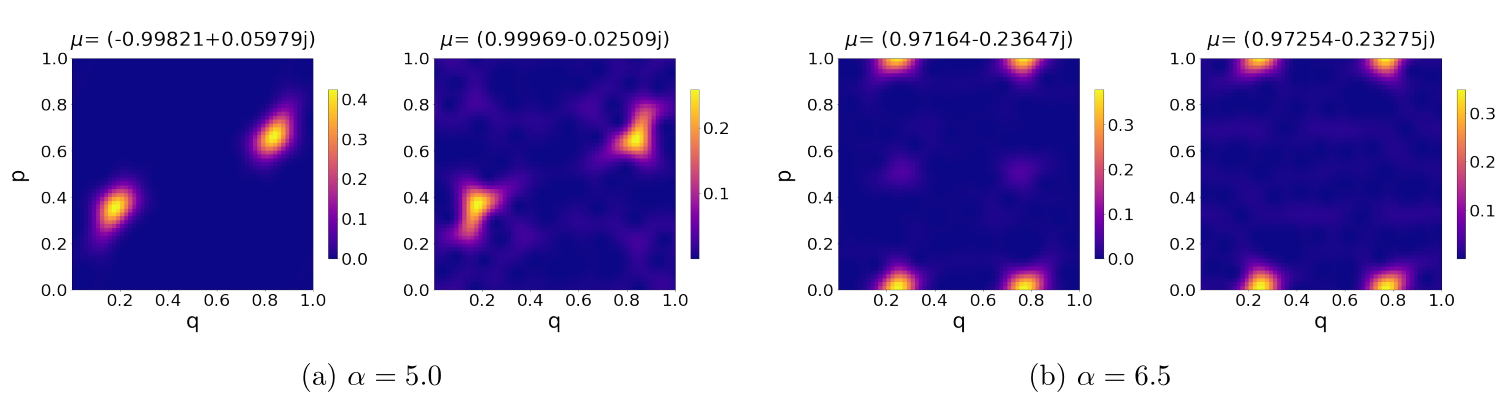}
\caption{\label{fig:mode_unit_submat} The Husimi function of the two eigenstates of the local unitary $U_{1}$ contributing the most to the largest mode of the submatrix $\mathcal{M}_{S}$, with $b=0.1$ for (a) $\alpha=5.0$, and  (b) $\alpha=6.5$, and $N=60$.}
\end{figure}

Due to the symmetry noted above for $d_{ij}$, the proper submatrix, $\mathcal{M}_{S}$ is Hermitian. It seems a peculiar situation where a truncation of a non-Hermitian matrix is Hermitian. Figure~({\ref{fig:eigval_submat_chan}}) shows the variation of the largest eigenvalue of the submatrix, $\mathcal{M}_{S}$ and the quantum channel, $\mathcal{M}(U)$ as we vary, $\alpha$ for $b=0.1$. This numerical data seems to indicate that the largest eigenvalue of $\mathcal{M}_{S}$ bounds the largest eigenvalue of $\mathcal{M}(U)$ from below. This will be the case for proper submatrices of Hermitian operators from the well-known eigenvalue interlacing theorems. Moreover for small values of $\alpha$, the Hermitian submatrix (only of size $N$) seems to capture very well the full channel's most significant mode.

The largest eigenmode of the submatrix, $\mathcal{M}_{S}$ in the free basis, $\{|\phi_{i}\kt |\phi_{i}^*\kt\}$ is of interest. We observe that in cases where there are short stable periodic orbits, even in small islands, they seem to have contribution only from the modes of $U_{1}$ corresponding to the eigenvalues that are nearly degenerate or $\pi$ apart, corresponding to fixed or period-2 points. The eigenmodes of the $U_{1}$ corresponding to these eigenvalues are the ones that carry the information about the periodic islands in the phase space. This is the origin of the selection that is observed for the localized modes of the channel.

The Husimi function of the eigenstates of the local unitary $U_{1}$ associated with these two eigenvalues are the ones that are localized on these stable periodic orbits at $\alpha =5.0$, Fig.~(\ref{fig:mode_unit_submat}). On the other hand, for $\alpha=6.5$, the modes that contribute the most to the largest eigenmode of $\mathcal{M}_{S}$ are associated with nearly degenerate eigenvalues, Fig.~(\ref{fig:mode_unit_submat}).
The difference between the two modes is that one is scarred by a period two orbit, and the other is scarred by a fixed point. Thus an analysis of the submatrix may form a route for a more comprehensive understanding of the channels, especially for small values of the interaction or noise parameter, $b$.

\section{\label{sec:Summary}Summary and discussion}

This work may be placed in the context of the current vigorous pursuit of quantum chaos in open systems with the major advent of quantum technologies. 
Starting from the classical Koopman operator, evolving functions in phase space, four channels have been derived which describes a subsystem evolution, with the ``environment" now being reduced to a stochastic element. This is exact and leads to subunitary classical Koopman channels whose spectrum and modes can be compared to the traditionally constructed quantum channels. In a way this inverts the quantum-classical correspondence and we find signatures of quantum chaos in the classical domain, which is now dominated by eigenvalues and an infinite number of modes. Numerically, these open classical channels are more accessible to truncation than the unitary ones. 

Constructing standard map channels in both the classical and quantum domains we compared the spectral features and found there to be remarkable correspondence in the semiclassical domain. At relatively low to medium noise levels, both also showed features that are easily correlated to the dynamical features of a single ``clean" standard map. The quantum channel spectra seem to fall into two sets, one that is in a dense annular region, and the others that are mostly exterior to it. This separation is more clear in the low interaction/noise and low chaos regimes. The annular region seems to obey the single-ring theorem of non-Hermitian RMT, although there are some deviations that need further study. This  regions shrinks and in the classical limit can vanish. Thus the states that survive the classical limit are perhaps the most interesting ones and are dominated by stable periodic orbits when they are present. However, rather than the orbits themselves, the unstable manifolds seem to dominate the eigenmodes. Parts of the unstable manifold ``light up" differently in these survivor eigenmodes. Thus these studies will likely contribute to a better understanding of quantum  scarring as well as chaos assisted tunneling.

In a recent paper \cite{Ergodic_channels_Aravinda}, the authors have studied the quantum channels considered in this paper, but for many-body systems after partitioning them into two halves. In particular they have shown eigenvalues of the channel for a short-range fermionic model and for the long-range SYK model. The clear differences between these models, with the SYK eigenvalues clustering more at the origin is notable. Also, just as in the cases studied here there are exceptional strings of states with real eigenvalues and one may speculate if these are localized or many-body scarred in some sense. Of course, this possibility opens up to the whole regime of many-body physics and the general channels that they describe. We have only skimmed the surface of a rich landscape, and there are many evident questions and directions. One, is the generalization to the two subsystems in the quantum case having unequal dimensions. The other is to have a state of the ``environment" that is not the most mixed. This leads to a weighting of the noise in both the classical and quantum channels. A semiclassical theory of these channels promises to be more benign in terms of convergence issues than the unitary one. It would be interesting to have a perturbative analysis for small interaction strengths especially, when the submatrix method mentioned here is seen to hold. The alleged split of states into those that survive the classical limit and those that do not, in the sense that their eigenvalues do not vanish needs to be more thoroughly investigated, both numerically and hopefully analytically. Another aspect that may be studied is the effect of symmetry breaking. It is very simple in these models to introduce phases that break parity symmetry or time-reversal, both of which are present in the cases presented here.

\begin{acknowledgments}
BV thanks MPhasis and CQuiCC, IIT Madras, for support through a fellowship.
\end{acknowledgments}

\bibliography{refer}

\end{document}